\documentclass[twocolumn]{autart}    

\usepackage{graphicx}          

\usepackage{amssymb,amsmath,graphicx,multicol,float,bm}

\usepackage[fleqn]{nccmath}
\usepackage{subcaption}
\usepackage[colorlinks]{hyperref}
\usepackage{epstopdf}
\usepackage{makecell}
\usepackage{enumitem}
\usepackage{wrapfig}
\usepackage[dvipsnames]{xcolor}
\graphicspath{{figs/}}
\usepackage[normalem]{ulem}

\newtheorem{Remark}{Remark}[section]
\newtheorem{Corollary}{Corollary}[section]
\newtheorem{Definition}{Definition}[section]
\newtheorem{Theorem}{Theorem}[section]
\newtheorem{Proposition}{Proposition}[section]
\newtheorem{proof}{Proof}
\newtheorem{Lemma}{Lemma}[section]
\newtheorem{Assumption}{Assumption}[section]

\newcommand{\BM}{\begin{matrix}}
\newcommand{\EM}{\end{matrix}}

\newcommand{\ep}{\epsilon}

\newcommand{\ba}{\begin{array}}
\newcommand{\ea}{\end{array}}
\newcommand{\be}{\begin{eqnarray}}
\newcommand{\ee}{\end{eqnarray}}
\newcommand{\EQQ}{\begin{eqnarray*}}
\newcommand{\ENN}{\end{eqnarray*}}

\def\tw{{\textup w}}
\def\ttpl{{\textup {Tpl}}}

\def\tt#1#2{\left[\begin{array}{c}#1\\ #2\end{array}\right]}

\newcommand{\bd}{\begin{Definition}
\begin{rm} }
\newcommand{\ed}{ \end{rm}
\end{Definition} }

\newcommand{\bexercise}{\begin{exercise}\vspace{-3mm}
\begin{rm} }
\newcommand{\eexercise}{ \end{rm}
\end{exercise} }

\newcommand{\bappend}{\begin{append}\vspace{-3mm}
\begin{rm} }
\newcommand{\eappend}{ \end{rm}
\end{append} }

\newcommand{\basm}{\begin{Assumption} \begin{rm}}
\newcommand{\easm}{\end{rm} \end{Assumption}}

\newcommand{\bpropty}{\begin{property} \vspace{-3mm}\begin{rm}}
\newcommand{\epropty}{\end{rm} \end{property}}

\newcommand{\bremark}{\begin{Remark}
\begin{rm} }
\newcommand{\eremark}{\hfill$\lozenge$ \end{rm}\end{Remark} }
\newcommand{\bt}{\begin{Theorem} \begin{rm} }
\newcommand{\et}{ \end{rm}
\end{Theorem} }
\newcommand{\bl}{\begin{Lemma} \begin{rm} }
\newcommand{\el}{ \end{rm}
\end{Lemma} }
\newcommand{\bcorollary}{\begin{Corollary} \begin{rm} }
\newcommand{\ecorollary}{ \end{rm}
\end{Corollary} }
\newcommand{\bdefinition}{\begin{Definition}\begin{rm} }
\newcommand{\edefinition}{ \end{rm}
\end{Definition} }
\newcommand{\bproposition}{\begin{proposition} \begin{rm} }
\newcommand{\eproposition}{ \end{rm}
\end{proposition} }
\newcommand{\bexample}{\begin{example} \begin{rm} }
\newcommand{\eexample}{ \end{rm}
\end{example} }

\newcommand{\bproof}{\begin{proof} \begin{rm} }
\newcommand{\eproof}{ \end{rm} \end{proof} }

\allowdisplaybreaks[4]

\begin{document}

\begin{frontmatter}

\title{On Kernel Design for Regularized Volterra Series Identification of Wiener-Hammerstein Systems}

\thanks[footnoteinfo]{Corresponding author Tianshi Chen.
\\This work was supported by NSFC under
Grant 62273287 and the Shenzhen Science and Technology Innovation Council under Grant JCYJ20220530143418040 and Grant
JCY20170411102101881.
}

\vspace{-7mm}
\author[1]{Yu Xu},
\author[2]{Biqiang Mu}
and
\author[1]{Tianshi Chen}

\address[1]{School of Data Science, The Chinese University of Hong Kong, Shenzhen, 518172, China, (e-mail:yuxu19@link.cuhk.edu.cn;
tschen@cuhk.edu.cn).}

\address[2]{Key Laboratory of Systems and Control, Institute of Systems Science, Academy of Mathematics and Systems Science, Chinese Academy of Sciences, Beijing 100190, China, (e-mail:bqmu@amss.ac.cn).}

\vspace{-4mm}
\begin{abstract}                
There have been increasing interests on the Volterra series identification with the kernel-based regularization method. The major difficulties are on the kernel design and efficiency of the corresponding implementation. In this paper, we first assume that the underlying system to be identified is the Wiener-Hammerstein (WH) system with polynomial nonlinearity. We then show how to design kernels with nonzero off-diagonal blocks for Volterra maps by taking into account the prior knowledge of the linear blocks and the structure of WH systems. Moreover, exploring the structure of the designed kernels leads to the same computational complexity as the state-of-the-art result, i.e., $O(N^3)$, where $N$ is the sample size, but with a significant difference that the proposed kernels are designed in a direct and flexible way. In addition, for a special case of the kernel and a class of widely used input signals, further exploring the separable structure of the output kernel matrix can lower the computational complexity from $O(N^3)$ to $O(N\gamma^2)$, where $\gamma$ is the separability rank of the output kernel matrix and can be much smaller than $N$. We finally run Monte Carlo simulations to demonstrate the proposed kernels and the obtained theoretical results.
\end{abstract}

\vspace{-4mm}

\begin{keyword}
Wiener-Hammerstein systems, Volterra series identification, kernel-based regularization method, kernel design.
\end{keyword}

\end{frontmatter}


\section{Introduction}\label{sec:intro}

Volterra series is a popular model structure for nonlinear system identification due to its capability of representing many nonlinear systems, e.g., the nonlinear time-invariant systems with fading memory \cite{BC85}. A typical issue in the Volterra series identification is the exponential growth of the number of model parameters, denoted by $n_\theta$ below, with respect to (w.r.t.) both the order of the nonlinearity and the memory length. In order to obtain a Volterra series with good quality, a large number of data is usually required, resulting in heavy computational burden. If the data record is short and/or if the data has a low signal-to-noise ratio (SNR), the obtained model estimate is often subject to large variance. One recent trend to deal with this problem is by using the kernel-based regularization method (KRM). In the past decade, KRM for the linear time-invariant (LTI) system identification has been explored, e.g., the survey papers \cite{PDCDL14,CHIUSO16,LCM20} and the book \cite{PCCNL22}. It has the feature that it finds a systematic way to engage the prior knowledge of the underlying system to be identified in the system identification loop, particularly in the selection of both the model structure and the model complexity. The carrier of the prior knowledge is the so-called kernel, which determines the model structure, and whose hyper-parameter determines the model complexity.

KRM has been extended to the Volterra series identification, e.g., \cite{SWH17,BMLS17,BCS18,LCP21}, with major difficulties on the kernel design and efficiency of the corresponding implementation. The prior knowledge considered in \cite{SWH17,BMLS17,BCS18,LCP21} is the same, i.e., the smooth and exponentially decaying property of Volterra maps \footnote{\label{ftnt:VolterraMap}The terminology \textit{Volterra map} is traditionally called \textit{Volterra kernel} in the literature, but similar to \cite{LCP21}, we use \textit{Volterra map} to avoid confusion with the concept of \textit{kernel} from machine
learning.}, which are the multi-dimensional generalization of the impulse response of LTI systems involved in each term of the Volterra series. However, different ways were taken to design the kernels in \cite{SWH17,BMLS17,BCS18,LCP21}. In \cite{SWH17,BMLS17,BCS18}, the kernels were designed for the \emph{Volterra maps}, and the corresponding implementation has a computational complexity of $O(Nn_\theta^2)$, where $N$ is the sample size.  Since $n_\theta$ grows exponentially and can be far larger than $N$, KRM in \cite{SWH17,BMLS17,BCS18} is computationally very expensive except for nonlinear systems with both weak nonlinearity and short memory length. To reduce the computational complexity, the kernel in \cite{LCP21} was designed for the \emph{Volterra series}, and the corresponding implementation has a computational complexity of $O(N^3)$. The price paid is that the kernel in \cite{LCP21} can only embed the prior knowledge of the Volterra maps in an \emph{indirect} and \emph{rigid} way because of the unnecessary inclusion of the input signals in the kernel design. To understand this, simply note that the relation between the input signals and the Volterra maps is known exactly from the Volterra series and thus it is not necessary to design a kernel for both the input signals and the Volterra maps, that is, it is not necessary to design a kernel directly for the Volterra series. Indeed, such design may break down this relation, and clearly, if the designed kernel does not reflect this relation, nontrivial bias will be resulted. Besides, a common problem in \cite{SWH17,BMLS17,BCS18,LCP21} is that instead of specifying the underlying system to be identified, it was simply assumed that the Volterra maps are smooth and exponentially decaying, but it is unclear where this kind of prior knowledge is from. Their results are ``seemingly general" for the Volterra series identification, but in essence can only be applied to a class of nonlinear systems, which is not clear to the users. Then it is worth to stress that it is important to specify the underlying system to be identified, for at least two reasons: first, more precise and richer prior knowledge can be obtained from the underlying system and used in the kernel design; second,  when applying the designed kernel to a system that does not have the prior knowledge, nontrivial bias will be resulted. Another common problem in \cite{SWH17,BMLS17,BCS18,LCP21} is that  all kernels designed for or corresponding to the \emph{Volterra maps} in \cite{SWH17,BMLS17,BCS18,LCP21} have block-diagonal structure, but none makes use of the off-diagonal blocks.

In this paper, we study the issue of kernel design for the Volterra series identification and the corresponding efficient implementation. Different from \cite{SWH17,BMLS17,BCS18,LCP21}, we first assume that the underlying system to be identified is the Wiener-Hammerstein (WH) system with polynomial nonlinearity. There are several reasons to consider this class of nonlinear systems. First, their parallel interconnections have been proven to be able to approximate arbitrarily well nonlinear time-invariant systems with fading memory, \cite[Theorem 2]{BC85}, \cite{PALM79}, and thus they have been one of the most widely studied classes of nonlinear systems in system identification \cite{BAI10}, \cite[page 75]{SL19}. Second, since they can be represented by the Volterra series \cite{BAI10}, more precise and richer prior knowledge of Volterra maps can be obtained from that of them. Third, since the simulation examples in \cite{SWH17,BMLS17,BCS18,LCP21} all included WH systems or their special case, it is interesting to design kernels for them and make comparison between the designed kernels with the ones in \cite{SWH17,BMLS17,BCS18,LCP21}. We then show how to design kernels with nonzero off-diagonal blocks for \emph{Volterra maps} by taking into account the prior knowledge of the linear blocks and the structure of WH systems. Moreover, exploring the structure of the designed kernels leads to the same computational complexity as the one in \cite{LCP21}, i.e., $O(N^3)$, but with significant difference that the proposed kernels are designed to embed the prior knowledge of the \emph{Volterra maps} in a \emph{direct} and \emph{flexible} way. In addition, for a special case of the kernel and a class of widely used input signals, we further show that exploring the rank structure of the output kernel matrix can lower the computational complexity from $O(N^3)$ to $O(N\gamma^2)$, where $\gamma$ is the separability rank of the output kernel matrix and can be much smaller than $N$. We finally run Monte Carlo simulations to demonstrate the proposed kernels and obtained theoretical results.


Finally, since the identification of WH systems has been studied extensively, it is natural to question the advantage of the proposed approach over the existing ones. Indeed,  there are different streams of approaches for the identification of WH systems, which differ in the choice of the model structure, the estimation method, and etc. For example, \emph{one stream of approaches} is the classical maximum likelihood/prediction error (ML/PE) parametric methods, e.g., \cite{EL05,HLW08,TS14,SK17}. Parametric model structures are postulated for both the linear and nonlinear blocks, and their model complexities are governed by the number of parameters, of \emph{integer values}, and tuned by cross validation or model structure selection criteria such as Akaike’s information criterion (AIC), Bayesian information criterion (BIC), and etc., which are however not as reliable as expected especially when the data is short and/or has a low SNR, see e.g., \cite{COL12,PDCDL14,PCCNL22}. In contrast, the proposed approach is a non-parametric approach, and the kernel plays a key role: its structure determines the underlying model structure of the Volterra series, and its model complexity is governed by the hyper-parameter, of \emph{real values}, and tuned by e.g., the empirical Bayes method, which is known to be robust when the data is short and/or has a low SNR, see e.g., \cite{PC15,MCL18,JMLC23}. \emph{Another stream} is the Bayesian semi-parametric or non-parametric methods, e.g., \cite{PC09,LSJ13,RBH17,RLH19}. Gaussian processes (GPs) are used to model the nonlinear block (semi-parametric) \cite{LSJ13} or both the linear and nonlinear blocks (non-parametric) \cite{PC09,RBH17,RLH19}. Prior knowledge of the linear and/or nonlinear blocks is engaged in the system identification loop through a well designed kernel, i.e., the covariance function of the GP. However, since the posterior is analytically intractable, different kinds of approximation methods are proposed in \cite{PC09,LSJ13,RBH17,RLH19} to  approximate either the posterior or the posterior mean. In contrast, since the over-parameterization of the WH system as a Volterra series leads to a linear regression problem, the proposed non-parametric approach is analytically tractable and thus the computation is exact and without any approximation. We have also run Monte Carlo simulations to compare the proposed approach with the ones in \cite{HLW08,LSJ13}, and our simulation results show that the proposed approach can lead to better average accuracy and robustness.

The remaining parts of this paper are organized as follows. In Section 2, we introduce the preliminary and problem statement. In Section 3, we focus on the kernel design, and in Section 4, we discuss the computational aspects. In Section 5, we run numerical simulations to show the efficacy of the designed kernels. Finally, we conclude the paper in Section 6. All proofs of theorems, corollaries and propositions are included in Appendix \ref{Appendix}.

\section{Preliminary and Problem Statement}
\label{sec:probstat}
In this paper, we consider the identification of a single-input single-output finite Volterra series model, e.g., \cite{WIENER58,SCHETZEN80} and \cite[Eq. (S68)]{SL19}, as follows:
\begin{align}\label{eq:VolterraSeriesModel}
 y(t) &= f(\bm{u}_t)+v(t),~t=1,\ldots,N,  \\
       f(\bm{u}_t)&\triangleq h_0+\sum_{m=1}^M\sum_{\tau_1=0}^{ n-1}\cdots \sum_{\tau_m=0}^{ n-1}h_m(\bm{\tau}_m)\prod_{\tau=\tau_1}^{\tau_m}u(t-\tau), \notag
  \end{align}
where $t=1,\ldots,N$ are the time indices, $N\in\mathbb{N}$ is the sample size, $u(t)\in\mathbb{R}$, $y(t)\in\mathbb{R}$, $v(t)\in\mathbb{R}$ are the input, measurement output, and measurement noise at time $t$, respectively, $v(t)$ is an i.i.d. Gaussian noise with zero mean and variance $\sigma^2>0$, $\bm{u}_t=[u(t),u(t-1),\cdots,u(t-n+1)]^T$, $M\in\mathbb{N}$ is the order of the Volterra series $f(\bm{u}_t)$, $n\in\mathbb{N}$ is the memory length of the Volterra series $f(\bm{u}_t)$, the constant $h_0\in\mathbb{R}$ is the $0$-th order Volterra map, and for $m=1,\ldots,M$, $h_m(\bm{t}_m)\triangleq h_m(t_1,\ldots,t_m)$ with $\bm{t}_m=[t_1,\cdots,t_m]^T$ is the $m$-th order Volterra map. The goal of Volterra series identification is to estimate the Volterra map coefficients $h_m(\bm{t}_m)$, $t_i=0,\ldots,n-1$, $i=1,\ldots,m$, $m=1,\ldots,M$, as well as possible based on the input-output data $y(t),u(t)$, $t=1,\ldots,N$ with $n\leq N$.
\begin{Remark}\label{rmk:truncation}
It is worth to note that the finite Volterra series model in \eqref{eq:VolterraSeriesModel} is a truncation of the infinite one \cite{BC85}, i.e.,
\begin{subequations}
    \begin{align}
      \begin{split}
f(\bm{u}_t) &=h_0+  \sum_{m=1}^M\sum_{\tau_1=0}^{ \infty}\cdots \sum_{\tau_m=0}^{\infty}h_m(\bm{\tau}_m)\prod_{\tau=\tau_1}^{\tau_m}u(t-\tau),
\end{split}\\
\begin{split}\label{eq:conv_condition}
&\sum_{t_1=0}^{ \infty}\cdots \sum_{t_m=0}^{\infty}\left |h_m(\bm{t}_m)\right|<\infty,
\end{split}
\end{align}
\end{subequations}
with $\bm{u}_t=[u(t),u(t-1),\ldots]^T$ and this truncation is not a limitation since \eqref{eq:conv_condition} indicates $\lim_{t_i\to \infty} h_m(\bm{t}_m)=0, \forall i = 1,\ldots,m$. In practice, the memory length $n$ and the order of the Volterra series $M$ can be chosen sufficiently large such that the bias due to the truncation can be arbitrarily small.
\end{Remark}
\subsection{The kernel-based regularization method}\label{sec:krm}
The Volterra series model \eqref{eq:VolterraSeriesModel} can be rewritten as a linear regression model in the following matrix-vector format:
\begin{align}
\label{eq:VecForm}
  Y &= \Phi\theta+V,\ \Phi = [\Phi_0, \cdots, \Phi_M],
\theta = [\theta_0^T, \cdots, \theta_M^T]^T
\end{align}
where $Y=[y(1),\cdots,y(N)]^T$, $V=[v(1),\cdots,v(N)]^T$, $\Phi_0=[1,\cdots,1]^T\in\mathbb{R}^{N}$, $\theta_0=h_0$, and for $m=1,\ldots,M$, the $t$-th row of $\Phi_m\in\mathbb{R}^{N\times n^m}$ \footnote{There are unknown inputs in \eqref{eq:VecForm}, i.e., $u(0)$, $u(-1)$, $\ldots$, $u(-n+2)$ and there are two ways to handle them. The first way is to use the pre-windowing technique \cite[page 320-321]{LJUNG99} to deal with unknown inputs, i.e., to replace the unknown inputs by zeros. The second way is to form \eqref{eq:VecForm} only based on known data. In this paper we do not have preference on either way. If the latter is used,  for Wiener system and WH system, the elements in $Y$ and $V$ in \eqref{eq:VecForm} can start from $y(n)$, $v(n)$ and $y(2n)$, $v(2n)$, respectively and $\Phi$ in \eqref{eq:VecForm} can be formed accordingly. \label{ftnt:ini_cond}} collects all the $m$-th order monomials of the lagged inputs in $\bm{u}_t$ and $\theta_m\in\mathbb{R}^{n^m}$ collects the coefficients $h_m(t_1,\ldots,t_m)$.

The parameter $\theta$ in \eqref{eq:VecForm} and the predicted output at time $t$ can be estimated by the kernel-based regularization method as follows:
\begin{subequations}  \label{eq:RLSestimators}
    \begin{align}
         \begin{split} \label{eq:RLS}
      \hat{\theta}^{\text{R}} &= \arg \min_{\theta} \|Y-\Phi\theta\|^2_2+\sigma^2\theta^T P^{-1}\theta\\
      &=P \Phi^T(\Phi P \Phi^T+\sigma^2 I_{N})^{-1} Y,
  \end{split}\\
  \begin{split}\label{eq:RLSoutput}
      \hat{y}^{\text{R}}(t)&=\bm{\phi} \hat{\theta}^{\text{R}} =\bm{\phi} P \Phi^T(\Phi P \Phi^T+\sigma^2 I_{N})^{-1}Y,
  \end{split}
      \end{align}
\end{subequations}
where $I_{N}$ is the $N$-dimensional identity matrix, $\bm{\phi}=[1,\bm{\phi}_1,\cdots,\bm{\phi}_M]\in\mathbb{R}^{1\times n_\theta}$ with $\bm{\phi}_m\in\mathbb{R}^{1\times n^m}$ collecting all $m$-th order monomials of the lagged inputs in $\bm{u}_t$, $m=1,\ldots,M$, $P\in \mathbb{R}^{n_{\theta}\times n_{\theta}}$ is a positive semidefinite matrix\footnote{The case where $P$ is singular is discussed in \cite[Remark 1]{PDCDL14} and \cite[Remark 3.1]{PCCNL22}. In particular, if $P$ is singular,  $P^{-1}$ can be replaced by its pseudoinverse $P^\dagger$ in \eqref{eq:RLSestimators}, and $\hat{\theta}^{\text{R}}=P \Phi^T(\Phi P \Phi^T+\sigma^2 I_{N})^{-1} Y$ is still the optimal
solution of \eqref{eq:RLSestimators}.}, $n_\theta=1+\sum_{m=1}^Mn^m$. The $(i,j)$-th element of $P$ can be designed through a positive semidefinite kernel ${\kappa}(t,s;{\eta}):\mathbb{R}_+\times \mathbb{R}_+\to \mathbb{R}$ with $\eta\in\Omega\subset \mathbb{R}^{\ell}$ being the hyper-parameter and $\Omega$ being a set in which $\eta$ takes values,  and thus $P$ is often called the kernel matrix. The noise variance $\sigma^2$ and the hyper-parameter $\eta$ can be estimated by the empirical Bayes (EB) method:
\begin{align}\label{eq:GML}
    \{\hat{\eta},\,&\widehat{\sigma^2}\} = \arg\min_{\eta \in \Omega, \sigma^2>0} Y^T\left(Q(\eta)+\sigma^2 I_{N}\right)^{-1}Y \notag \\&+ \log\det\left(Q(\eta)+\sigma^2 I_{N}\right), Q(\eta)=\Phi P(\eta)\Phi^T
\end{align}
where $Q(\eta)$ is the \emph{output kernel matrix}, e.g., \cite{PDCDL14}.

We focus on two issues for the estimation of $\theta$ by using  \eqref{eq:RLSestimators} and \eqref{eq:GML}. The first one is the kernel design, that is to embed the prior knowledge of $\theta$ in a kernel $\kappa(t,s;\eta)$ by parameterizing it with the hyper-parameter $\eta$. For example, for the impulse response estimation of LTI systems, i.e., $M=1$ in \eqref{eq:VolterraSeriesModel}, the widely used diagonal correlated (DC) kernel was designed in \cite{COL12} and given by
  \begin{align}
    \label{eq:Kernels}
    &\kappa^{\text{DC}}(t,s;\eta)=c^2e^{-\alpha(t+s)}e^{-\beta|t-s|},\nonumber\\
    &\eta=[c,\alpha,\beta]^T\in\Omega=\{c\in\mathbb{R},\alpha\in\mathbb{R}_{++}, \beta\in\mathbb{R}_{+}\}.
\end{align}
This kernel encodes the prior knowledge of the impulse response of the LTI system to be identified. The second issue is that straightforward calculation of \eqref{eq:RLSoutput} and \eqref{eq:GML} has high computational complexity $O(Nn_{\theta}^2)$, e.g., \cite{SWH17,BMLS17,BCS18}, which is extremely demanding for large $M$, since $n_\theta$ grows exponentially w.r.t. $M$.

To address these issues, we extend the guideline of kernel design for impulse responses \cite{COL12,CHEN18} to Volterra maps in (\ref{eq:VolterraSeriesModel}). First,  we derive the optimal kernel matrix for the Volterra maps by \cite[Theorem 1]{COL12}.
\begin{Lemma}[{\cite[Theorem 1]{COL12}} ]\label{prop:OptimalKernel}
 Consider the Volterra series \eqref{eq:VolterraSeriesModel} and its matrix-vector format \eqref{eq:VecForm}. Let $\theta^0=[(\theta_0^0)^T,(\theta_1^0)^T,\cdots,(\theta_M^0)^T]^T$ be the true value of $\theta$ in \eqref{eq:VecForm}. For a given kernel matrix $P$, let    $MSE(\hat{\theta}^{\text{R}}(P)) = \mathbb{E}(\hat{\theta}^{\text{R}}-\theta^{0})(\hat{\theta}^{\text{R}}-\theta^{0})^T$ be the mean square error  matrix of the regularized estimate $\hat{\theta}^{\text{R}}$ of $\theta$ defined in \eqref{eq:RLSestimators}, where  $\mathbb{E}$ represents the mathematical expectation. Then we have
  \begin{equation*}
          MSE(\hat{\theta}^{\text{R}}(P))\succeq MSE(\hat{\theta}^{\text{R}}(P^{\text{opt}})),~P^{\text{opt}}=\theta^{0}(\theta^{0})^T,
  \end{equation*}
where $P^{\text{opt}}$ is  the optimal kernel matrix, in the sense that $MSE(\hat{\theta}^{\text{R}}(P))- MSE(\hat{\theta}^{\text{R}}(P^{\text{opt}}))$ is positive semidefinite.
\end{Lemma}
Clearly, $P^{\text{opt}}$ cannot be applied in practice. 
However, it motivates a general guideline to design $P$ in  \eqref{eq:RLSestimators} in two aspects. Firstly, we should let the $\{p,q\}$-th block $P_{pq}$ of $P$ mimic the behavior of the $\{p,q\}$-th block of $P^{\text{opt}}$, i.e., $\theta^{0}_p(\theta^{0}_q)^T$, and the prior knowledge of $\theta^0_{p}$ and $\theta^0_{q}$ should be used in the design of $P_{pq}$. Secondly, besides embedding the prior knowledge, the kernel should be designed such that it can ease the computation of \eqref{eq:RLSestimators} and \eqref{eq:GML} if possible.

\subsection{Wiener-Hammerstein systems}
\label{subsec:WHsys}
Then one may wonder what prior knowledge is available for the Volterra maps. Clearly, it depends on the true Volterra series \eqref{eq:VolterraSeriesModel} to be identified.
Then it is interesting to note from e.g., \cite{BAI10} that some block-oriented nonlinear systems, e.g.,  Wiener-Hammerstein (WH) systems, can be represented by \eqref{eq:VolterraSeriesModel} and in such cases,  the prior knowledge of the Volterra maps can be obtained from that of the WH systems and used in the kernel design.

\begin{figure}
  \centering
        \includegraphics[scale=0.58]{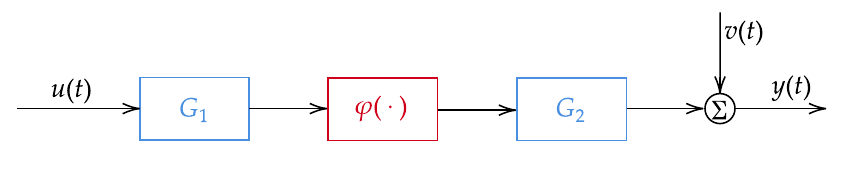}
        \vspace{-4mm}
        \caption{The block diagram of the WH system where $G_1$ and $G_2$ are the linear blocks and $\varphi(\cdot)$ is the static nonlinearity.}
        \label{fig:WienerHammersteinSystem}
\end{figure}

To be specific, we recall the WH systems as shown in Fig. \ref{fig:WienerHammersteinSystem}, where $G_1$ and $G_2$ are LTI systems, called \emph{the linear blocks} and $\varphi(\cdot)$ is a memoryless nonlinear function, called \emph{the static nonlinearity}. Here, $\varphi(\cdot)$ is assumed to be a polynomial of order $M$, e.g., \cite{SPE08,BAI10}, that is,
      \begin{align}
        \label{eq:StaticNonlinearity_WienerHammerstein}
        \varphi(z) = \sum_{m=0}^M a_mz^{m},~ a_0,\ldots,a_M\in\mathbb{R}.
      \end{align}

As is well known from e.g., \cite{BAI10}, assuming that the linear blocks are finite impulse response (FIR) models with order $n$, the WH system  can then be written in the form of the Volterra series \eqref{eq:VolterraSeriesModel} with
\begin{align}\label{eq:VolterraKernel_WienerHammerstein}
            h_m(t_1,\ldots,t_m)=&a_m \sum_{\tau=0}^{n-1}g_2(\tau)\prod_{i=1}^{m}g_1(t_i-\tau),
        \end{align}
      where $m=1,\ldots,M$, and $g_1(t)$ and $g_2(t)$ are the impulse responses of the LTI systems $G_1$ and $G_2$, respectively.
Clearly, if the linear blocks $G_1$ and $G_2$ are known to be bounded-input bounded-output stable (simply called stable below), then $g_1(t)$ and $g_2(t)$ will be absolutely summable. Moreover, the convolution between $g_2(t)$ and $\prod_{i=1}^{m}g_1(t_i)$ in \eqref{eq:VolterraKernel_WienerHammerstein} reflects the prior knowledge of the structure of the WH system. These prior knowledge can then be used in the kernel design for the Volterra maps.
\begin{Remark}\label{rmk:poly_nonlinearity}
It is worth to note that the WH system belongs to the class of nonlinear time-invariant system with fading memory, which can be approximated arbitrarily well by the parallel interconnections of WH systems with polynomial nonlinearities \cite[Theorem 2]{BC85}, \cite{PALM79}. The problem is that the polynomials are unknown and therefore if the orders of the polynomial nonlinearities are chosen sufficiently large such that the model contains the true system as a special case, the bias due to truncation of the order of the polynomial can be negligible.
\end{Remark}

\begin{Remark}\label{rmk:PE}
The excitation condition for the identification of Volterra series has been studied before. For example, it was shown in \cite{NV94} that a white Gaussian noise (i.i.d. Gaussian distributed with zero mean and unit variance) is persistently exciting for any finite order $M$ and any finite memory length $n$. The design of more general input signal was discussed in \cite[Chapter 7]{Birpoutsoukis2018}, and in particular, the classical $D$-optimal input design \cite{Wong1994} problem was studied with two different constraints. The first one imposes that the input signal is bounded by a given threshold, and the second one imposes that the input signal contains the deterministic time sequences, whose amplitude can only take values from a finite, predefined set of numbers. Interested readers are referred to \cite{NV94}, \cite[Chapter 7]{Birpoutsoukis2018} for more detailed discussions.
\end{Remark}

\subsection{Problem statement} \label{subsec:problem_statement}
As mentioned in Section \ref{sec:intro}, KRM has been extended to the Volterra series identification, e.g., \cite{SWH17,BMLS17,BCS18,LCP21}, with major difficulties on the kernel design and efficiency of the corresponding implementation. The attempts in \cite{SWH17,BMLS17,BCS18,LCP21} have shown its great potential, but there are also some issues that need to be addressed to further release its potential.  In this paper, we aim to address these issues. Firstly,
we consider the Volterra series model \eqref{eq:VolterraSeriesModel} under the assumption that the true system is the WH system with the polynomial \eqref{eq:StaticNonlinearity_WienerHammerstein} as the static nonlinearity and our goal is to design kernels between the Volterra maps $h_p(t_1,\ldots,t_p)$ and  $h_q(s_1,\ldots,s_q)$, $p,q=1,\ldots,M$, that is to embed the prior knowledge of both the structure and the linear blocks of WH systems in the kernels $\mathcal{K}_{pq}(\bm{t}_p,\bm{s}_q;\eta_{pq})$ by parameterizing it with some hyper-parameters $\eta_{pq}$. Secondly, we show that the designed kernels can lower the computational complexity for calculating  \eqref{eq:RLSoutput} and \eqref{eq:GML} from $O(Nn_\theta^2)$ to $O(N^3)$, and moreover, for a special case of the designed kernel and a class of widely used input signals, by exploring the separable structure of the output kernel matrix $Q$ in \eqref{eq:GML}, the computational complexity is further reduced to $O(N\gamma^2)$, where $\gamma$ is the separability rank of $Q$, irrespective of and can be much smaller than $N$.

\section{Kernel Design}\label{sec:kerneldesign}
To design kernels for the Volterra maps $h_m(t_1,\ldots,t_m)$, $m=1,\ldots,M$, in \eqref{eq:VolterraKernel_WienerHammerstein}, following from \cite[Theorem 1]{COL12}, we first derive the optimal kernel for the Volterra maps \eqref{eq:VolterraKernel_WienerHammerstein}.
\begin{Proposition}   \label{prop:OptimalKernel_offdiagonal_WienerHammersteinType}
 Consider the Volterra series \eqref{eq:VolterraSeriesModel} with \eqref{eq:VolterraKernel_WienerHammerstein}. Suppose that the true Volterra maps are
 $$h_m^{0}(\bm{t}_m)=a_m^0 \sum_{ \tau=0}^{n-1}g_2^0(\tau)\prod_{i=1}^{m}g_1^0(t_i-\tau), m=1,\ldots,M,$$ where $\bm{t}_m=[t_1,t_2,\cdots,t_m]^T$ with $m\in\mathbb N$, $a_m^0$ is the true value of $a_m$ in the static nonlinearity \eqref{eq:StaticNonlinearity_WienerHammerstein}, $g_1^0(t)$ and $g_2^0(t)$ are the true impulse responses of the linear blocks $G_1$ and $G_2$, respectively. Then for $p,q=1,\ldots,M$, the optimal kernel $\mathcal{K}_{pq}^{\text{opt}}(\bm{t}_p,\bm{s}_q)$ between $h_p(\bm{t}_p)$ and $h_q(\bm{s}_q)$ with $\bm{t}_p=[t_1,t_2,\cdots,t_p]^T$ and $\bm{s}_q=[s_1,s_2,\cdots,s_q]^T$ is
\begin{align} \nonumber
     &\mathcal{K}_{pq}^{\text{opt}}(\bm{t}_p,\bm{s}_q)=\sum_{\xi_1=0}^{n-1}\sum_{\xi_2=0}^{n-1}\kappa_2^{\text{opt}}(\xi_1,\xi_2)\mathcal{K}^{{\tw},\text{opt}}_{pq}(\bm{t}_p-\xi_1,\bm{s}_q-\xi_2),\\
        \label{eq:OptimalKernel_offdiagonal_WienerHammersteinType}
       &         \mathcal{K}^{{\tw},\text{opt}}_{pq}(\bm{t}_p,\bm{s}_q) =a_p^0a_q^0\\
         &\left\{
   \begin{aligned}\nonumber
   \prod_{i=1}^p\kappa_1^{\text{opt}}(t_i,s_i)\prod_{i=p+1}^qg_1^0(s_i)& , & p\leq q, \\
   \prod_{i=1}^q\kappa_1^{\text{opt}}(t_i,s_i)\prod_{i=q+1}^pg_1^0(t_i)& , & p> q,
   \end{aligned}
   \right.
       \end{align}
where $\kappa^{\text{opt}}_1(t,s)=g_1^0(t)g_1^0(s)$ and $\kappa^{\text{opt}}_2(t,s)=g_2^0(t)g_2^0(s)$ are the optimal kernels for the impulse responses of $G_1$ and $G_2$, respectively.
 \end{Proposition}

\begin{Remark}
\label{rmk:optimal_kernel}
The definition of optimal kernel is similar to that in \cite[Proposition 19]{PDCDL14}. That is, given the true Volterra map $h_p^0(\bm{t}_p)$, the optimal kernel is constructed by $\mathcal{K}_{pq}^{\text{opt}}(\bm{t}_p,\bm{s}_q)=h_p^0(\bm{t}_p)h_q^0(\bm{s}_q)$  for $p,q=0,1,\ldots,M$. Let $\theta_p^0$ be the vector that contains samples of $h_p^0(\bm{t}_p)$ with $\bm{t}_p=[t_1,\cdots,t_p]$ at $t_1,\ldots,t_p=0,\ldots,n-1$ and then form $\theta^0=[(\theta_0^0)^T,(\theta_1^0)^T,\cdots,(\theta_M^0)^T]^T$. The optimal kernel is defined through the optimal kernel matrix in Lemma \ref{prop:OptimalKernel}.
\end{Remark}

Following the guideline mentioned under \cite[Theorem 1]{COL12}, we propose to design the kernel for the Volterra maps \eqref{eq:VolterraKernel_WienerHammerstein} in two steps as follows:
\begin{enumerate}
\item[1)] use the prior knowledge of  $G_1$ and $G_2$ to design kernels $\kappa_1(t,s;\eta_1)$ and $\kappa_2(t,s;\eta_2)$ for the impulse responses $g_1(t)$ and $g_2(t)$, respectively, and design a function $\zeta(t;\eta_1)$ for the impulse responses $g_1(t)$;
\item[2)] replace $\kappa_1^{\text{opt}}(t,s)$, $\kappa_2^{\text{opt}}(t,s)$ and $g_1^0(t)$ by  $\kappa_1(t,s;\eta_1)$, $\kappa_2(t,s;\eta_2)$ and  $\zeta(t;\eta_1)$, respectively,
\end{enumerate}
and thus for $p,q=1,\ldots,M$, the kernel $\mathcal{K}_{pq}(\bm{t}_p,\bm{s}_q;\eta_{pq})$ between $h_p(\bm{t}_p)$ and $h_q(\bm{s}_q)$ takes the form of
\begin{subequations}
  \label{eq:off_diag_kernel_WienerHammersteinType}
\begin{align}\label{eq:off_diag_kernel_WienerHammersteinType_general}
      &\mathcal{K}_{pq}(\bm{t}_p,\bm{s}_q;\eta_{pq})=\\&\sum_{\xi_1=0}^{n-1}\sum_{\xi_2=0}^{n-1}\kappa_2(\xi_1,\xi_2;\eta_2)\mathcal{K}^{\tw}_{pq}(\bm{t}_p-\xi_1,\bm{s}_q-\xi_2;a_p,a_q,\eta_1),\nonumber
        \end{align}
      where $\eta_{pq}=[a_p,a_q,\eta_1^T,\eta_2^T]^T$ is the hyper-parameter and
      \begin{equation}
        \label{eq:off_diag_kernel_WienerType}
        \begin{split}
          \mathcal{K}^{\tw}_{pq}&(\bm{t}_p,\bm{s}_q;a_p,a_q,\eta_1) =a_pa_q\\
          &\left\{
    \begin{aligned}
    \prod_{i=1}^p\kappa_1(t_i,s_i;\eta_1)\prod_{i=p+1}^q\zeta(s_i;\eta_1)& , & p\leq q, \\
    \prod_{i=1}^q\kappa_1(t_i,s_i;\eta_1)\prod_{i=q+1}^p\zeta(t_i;\eta_1)& , & p> q.
    \end{aligned}
    \right.
        \end{split}
          \end{equation}
        \end{subequations}
Besides encoding the prior knowledge of $G_1$ and $G_2$, we also need to ensure the positive semidefiniteness of $P$ in \eqref{eq:RLSestimators} defined through the kernel \eqref{eq:off_diag_kernel_WienerHammersteinType}.

\begin{Theorem}\label{Prop:KernelPSD}
  The kernel matrix $P$ defined in \eqref{eq:RLSestimators} through the kernel \eqref{eq:off_diag_kernel_WienerHammersteinType} is positive semidefinite, if $\kappa_1(t,s;\eta_1)-\zeta(t;\eta_1)\zeta(s;\eta_1)$ is positive semidefinite.
\end{Theorem}

If the linear blocks $G_1$ and $G_2$ are both stable and overdamped, then we can design both $\kappa_1(t,s)$ and $\kappa_2(t,s)$ as the DC kernels in \eqref{eq:Kernels}. Moreover, we propose in this case to design $\zeta(t)$ in two ways in the following corollary.
\begin{Corollary}\label{prop:PSDzeta}
  Let $\kappa_1(t,s)=c_1^2e^{-\alpha_1(t+s)}e^{-\beta_1|t-s|}$, $\kappa_2(t,s)=c_2^2e^{-\alpha_2(t+s)}e^{-\beta_2|t-s|}$ and
  \begin{itemize}
    \item[1)] $\zeta(t)=c_1e^{-(\alpha_1+\beta_1)t}$,
    \item[2)] $\zeta(t)=c_1\big(\sum_{i=1}^{l}\sqrt{2}\ep_i\psi_i(t)\big)$ where $l\in\mathbb{N}$ is the number of bases chosen by the user, $l\ll N$, $\{\ep_i\}_{i=1}^l$ and $\{\psi_i\}_{i=1}^l$ are the eigenvalues and the orthonormal eigenfunctions of the DC kernel $\kappa_1(t,s)$ \cite{CHEN18TAC}, respectively. Specifically, $\ep_i=\frac{1}{(i-\frac{1}{2})^2\pi^2}$ and $\phi_i(t)=\sqrt{2}e^{(-\alpha_1+\beta_1)t}\sin\left((i-\frac{1}{2})\pi e^{-2\beta_1 t}\right)$.
  \end{itemize}
  Then in both cases, $P$ defined in \eqref{eq:RLSestimators} through the kernel \eqref{eq:off_diag_kernel_WienerHammersteinType} is positive semidefinite.
\end{Corollary}

\begin{Remark}\label{rmk:hyperparameters}
When the kernels $\kappa_1(t, s)$ and $\kappa_2(t, s)$ are chosen to be the DC kernels \eqref{eq:Kernels}, the hyper-parameters then include the $0$-th order Volterra map $h_0$, the polynomial coefficients $a_m$, $m=1,\ldots,M$, the noise variance $\sigma^2$ and the hyper-parameters  $c_1$, $c_2$, $\alpha_1$, $\alpha_2$, $\beta_1$, $\beta_2$ used to parameterize $\kappa_1(t, s)$ and $\kappa_2(t, s)$.
\end{Remark}

\section{Computational Aspects}\label{sec:Implementation}
The bottleneck for calculating \eqref{eq:RLSoutput} and \eqref{eq:GML} lies in the construction of the output kernel matrix $Q$, whose computational complexity is $O(Nn^2_{\theta})$ and $n_{\theta}$ is often far larger than $N$. In what follows, we show that there is an efficient way to construct $Q$ with computational complexity $O(N^3)$. For simplicity, we skip the dependence of the functions on the hyper-parameters from now on, e.g., we write $\mathcal{K}_{pq}(\bm{t}_p,\bm{s}_q;\eta_{pq})$ simply as $\mathcal{K}_{pq}(\bm{t}_p,\bm{s}_q)$.

\subsection{Efficient construction of the output kernel matrix}
The output kernel matrix can be efficiently constructed as stated in the following theorem.
\begin{Theorem}\label{Thm:OutputKernel}
Consider the kernel \eqref{eq:off_diag_kernel_WienerHammersteinType}. Let $K_1\in\mathbb{R}^{n\times n}$ and $K_2\in\mathbb{R}^{n\times n}$ be the kernel matrices defined by $[K_1]_{ij}=\kappa_1(i-1,j-1)$ and $[K_2]_{ij}=\kappa_2(i-1,j-1)$, respectively. Then the output kernel matrix $Q\in\mathbb{R}^{N\times N}$ can be calculated with computational complexity $O(N^3)$ by
\begin{equation}
  \label{eq:2dconv}
  Q = conv2(K_2,Q^{\tw}),
\end{equation}
where $conv2(\cdot,\cdot)$ denotes the 2-dimensional convolution\footnote{For $A\in\mathbb{R}^{n_1\times n_1}$ and $B\in\mathbb{R}^{n_2\times n_2}$ with $n_1,n_2\in\mathbb{N}$, the 2-dimensional convolution of $A$ and $B$ is a matrix whose $(k,\ell)$-th element is defined by
$
\sum_{i=1}^{\infty}\sum_{j=1}^{\infty}\mathcal{A}(i,j)\mathcal{B}(k-i,\ell-j).
$ where $\mathcal{A}(i,j)=[A]_{i,j}$ if $i,j=1,\ldots,n_1$ and zero otherwise, and $\mathcal{B}(i,j)=[B]_{i,j}$ if $i,j=1,\ldots,n_2$ and zero otherwise.} of two matrices and $Q^{\tw}\in\mathbb{R}^{N\times N}$ is defined as
\begin{align}\label{eq:OutputKernelMatrix_WienerSystem}
    Q^{\tw}
    &=\sum_{p>q}a_pa_q(\Psi K_1\Psi^T)^{\circ q} \circ \big(\Xi^{\circ p-q}\big) \notag\\&+ \sum_{p\leq q}a_pa_q(\Psi K_1\Psi^T)^{\circ p} \circ \big(\Xi^{\circ q-p}\big)^T,\\
\label{eq:Psi}
      \Psi &= \ttpl(\bm{u}_c,\bm{u}_r),
\end{align}
 where $\Psi = \ttpl(\bm{u}_c,\bm{u}_r)\in\mathbb{R}^{N\times n}$ is the Toeplitz matrix formed by its first column $\bm{u}_c\triangleq[u(1), u(2),\cdots,u(N)]^T\in\mathbb{R}^{N}$ and its first row $\bm{u}_r\triangleq[u(1), u(0),\cdots,u(-n+2)]\in\mathbb{R}^{1\times n}$ \footnote{ There are unknown inputs in $\Psi$ defined by \eqref{eq:Psi}, i.e., $u(0)$, $u(-1)$, $\ldots$, $u(-n+2)$ and there are two ways to handle them, as mentioned in footnote \footref{ftnt:ini_cond}. If we form \eqref{eq:VecForm} only with known data, then $\Psi$ can also be formed accordingly. \label{ftnt:ini_cond2}}, $A\circ B$ denotes the Hadamard product for two matrices $A$ and $B$ of the same size, for $k\in\mathbb{N}$, $A^{\circ k}$ denotes the $k$-th Hadamard power of $A$, $\Xi=\Psi\bm{\zeta}\mathbf{e}^T$ with $\bm{\zeta}=[\zeta(0), \zeta(1),\cdots,\zeta(n-1)]^T\in\mathbb{R}^{n}$ and $\mathbf{e}=[1,\cdots,1]^T\in\mathbb{R}^{N}$.
Moreover, \eqref{eq:RLSoutput} and \eqref{eq:GML} can be calculated with computational complexity $O(N^3)$ that is irrespective of $n_\theta$.
\end{Theorem}

\begin{Remark}\label{rmk:kerneldesign_volterramap_vs_series}
Theorem \ref{Thm:OutputKernel} places the proposed kernel \eqref{eq:off_diag_kernel_WienerHammersteinType} at the same level as the ones in \cite{LCP21} since they both have the computational complexity $O(N^3)$ of computing  \eqref{eq:RLSoutput} and \eqref{eq:GML}. However, they have a significant difference: the former is designed for the Volterra maps while the latter are designed for the Volterra series. It is worth to stress that  the former is better at encoding the specific prior knowledge of the Volterra maps, e.g., the convolution structure in \eqref{eq:off_diag_kernel_WienerHammersteinType} corresponding to the prior knowledge of the structure of the WH system.
\end{Remark}

\subsection{A special case: separable structure of the output kernel matrix}
\label{sec:RankStructure}

Now, we consider a special case of the kernel \eqref{eq:off_diag_kernel_WienerHammersteinType} with
\begin{align}\label{eq:k2sp}
  \kappa_2(t,s) = \left\{
\begin{aligned}
&1, &t=s=0, \\
&0, &\text{otherwise}.
\end{aligned}
\right.
\end{align}
Clearly, the kernel \eqref{eq:off_diag_kernel_WienerHammersteinType} in this case reduces to \eqref{eq:off_diag_kernel_WienerType} and embeds the structural prior knowledge of the WH system without the linear block $G_2$, i.e., the Wiener system in Fig. \ref{fig:WienerHammersteinSystem}. Accordingly, the output kernel matrix $Q$ in \eqref{eq:2dconv} becomes $Q^{\tw}$ in \eqref{eq:OutputKernelMatrix_WienerSystem}, i.e., $Q=Q^{\tw}$.

In this case, we show below that $Q$ can have the separable structure, which can be used to calculate \eqref{eq:RLSoutput} and \eqref{eq:GML} in a more efficient way.
Before proceeding to the details, we make the following assumption on the input signal $u(t)$ that there exist $\pi_i$, $\rho_i:\mathbb{R}_+\rightarrow \mathbb{R}$, $i=1,\ldots,r$ with $r\in\mathbb{N}$ such that
\begin{equation}\label{eq:InputCondition}
    u(t-b)=\sum_{i=1}^r \pi_i(t)\rho_i(b).
\end{equation}
This assumption is mild and many frequently used
test input signals in automatic control satisfy this assumption, e.g., \cite{CA21}. Then we have the following result.

\begin{Theorem}\label{Thm:LowRankOutputKernelMatrix_Full}
  Consider the Volterra series \eqref{eq:VolterraSeriesModel} and the kernel \eqref{eq:off_diag_kernel_WienerHammersteinType}. Assume that the input signal $u(t)$ satisfies \eqref{eq:InputCondition} and the kernel $\kappa_2(t,s)$ satisfies \eqref{eq:k2sp}. Then the output kernel matrix $Q=Q^\tw$ is separable\footnote{A matrix $A\in\mathbb{R}^{n_1\times n_2}$ is said to be separable with separability rank $p\in\mathbb{N}$ if $A=UV^T$ where $U\in\mathbb{R}^{n_1\times p},V\in\mathbb{R}^{n_2\times p}$ are called the generators of $A$.} with separability rank $\gamma =\frac{(r+M-1)!}{(r-1)!M!}+2\sum_{m=1}^{M-1}\frac{(r+m-1)!}{(r-1)!m!}$, and moreover, the computational complexity of \eqref{eq:RLSoutput} and \eqref{eq:GML} is $O(N\gamma^2+n^2r)$. If in addition, $\kappa_1(t,s)$ is extended $p$-semiseparable\footnote{\cite{CA21} A kernel $\kappa(t,s):\mathbb{R}_+\times \mathbb{R}_+ \to \mathbb{R}$ is said to be extended-$p$ semiseparable, if there exists $\mu_i,\nu_i:\mathbb{R}_+\to \mathbb{R}$, $i=1,\ldots,p$, such that $\kappa(t,s)=\sum_{i=1}^p \mu_i(t)\nu_i(s)$ for $t\geq s$, and
   $\kappa(t,s)=\sum_{i=1}^p \mu_i(s)\nu_i(t)$ for $t< s$.}
 with $p\in\mathbb{N}$ and $p\ll N$, then the computational complexity of \eqref{eq:RLSoutput} and \eqref{eq:GML} is reduced to $O(N\gamma^2)$.
\end{Theorem}

\begin{Remark}\label{rmk:comparison_with_SED-MPK}
Note that $\gamma$ in Theorem \ref{Thm:LowRankOutputKernelMatrix_Full} is irrespective of and can be much smaller than $N$.
Some examples of $\gamma$ w.r.t. $M$ and $r$ are given in Table \ref{tbl:SpRank}. We observe that $\gamma$ can be small w.r.t. $N$ if $M$ and/or $r$ is small. In other words, efficient implementations can be achieved if the nonlinearity is weak and/or the input has a simple structure.
When \eqref{eq:k2sp} holds and only the main diagonal blocks are considered, i.e., kernel \eqref{eq:off_diag_kernel_WienerHammersteinType} with $p=q=m$, $m=1,\ldots,M$, the corresponding output kernel matrix $Q$ is equivalent to that of the special case of the output kernel SED-MPK \cite[Eq. (36)]{LCP21}, where the hyper-parameters $\alpha_p$ (resp. $\beta_p$) are set to be equal to $\alpha$ (resp. $\beta$). Moreover, the separability rank $\gamma$ can be further reduced to $\sum_{m=1}^M\frac{(r+m-1)!}{(r-1)!m!}$.
\end{Remark}

\begin{Remark}\label{rmk:Kronecker}
When \eqref{eq:k2sp} holds and only the main diagonal blocks are considered, it is also interesting to observe that the kernel \eqref{eq:off_diag_kernel_WienerHammersteinType} with $p=q=m$, $m=1,\ldots,M$ reduces to $\mathcal{K}_{mm}^{\tw} (\bm{t}_m,\bm{s}_m) = a_m^2 \prod_{i=1}^m \kappa_1(t_i,s_i)$ and the kernel matrix $P=\text{diag}\{a_1^2K_1,a_2^2K_1\otimes K_1,\ldots,a_M^2K_1\otimes\cdots \otimes K_1\}$
where $\otimes$ denotes the Kronecker product. Although the kernel matrix $P$ does not have the Kronecker structure, its main diagonal blocks are different  Kronecker powers of $K_1$, which have the Kronecker structure, see e.g., \cite{ZORZI22}.
\end{Remark}

\begin{table}[]
\begin{tabular}{llllll}
\Xhline{0.8pt}
 & $r=1$  & $r=2$   & $r=3$   & $r=4$   & $r=5$     \\ \Xhline{0.8pt}
$M=2$                 & 3(2)  & 7(5)   & 12(9)  & 18(14) & 25(20)      \\\hline
$M=3$                  & 5(3)  & 14(9)  & 28(19)  & 48(34)  & 75(55)      \\ \hline
$M=4$                  & 7(4) & 23(14)  & 53(34)  & 103(69) & 180(125)    \\ \hline
$M=5$                  & 9(5) & 34(20)  & 89(55) & 194(125) & 376(251)   \\ \hline
\end{tabular}
\caption{Examples of the separability rank $\gamma$ w.r.t. $M$ and $r$ of the output kernel matrix corresponding to \eqref{eq:off_diag_kernel_WienerType}. The numbers within and without the parentheses denote the cases without or with the off-diagonal blocks, respectively.}
\label{tbl:SpRank}
\end{table}

\section{Numerical Simulation}\label{sec:simulation}

We run numerical simulations to test the proposed kernels and justify the obtained theoretical results.

\subsection{Test systems and databanks}
\label{subsec:systems_and_databanks}
In what follows, we generate 4 databanks D1-D4, associated with different types of test systems. We first introduce the test systems and then generate the databanks that contain input-output data from the test systems. First, we generate the test systems for D1-D4 as follows.
\begin{itemize}
    \item Databank D1 contains the nonlinear system in \cite[Fig. 3]{BMLS17}, which is shown in Fig. \ref{fig:exp1}. Specifically, the configuration is as follows: $G_0 = 2$, $G_1(q) = \frac{0.7568q^{-1}}{1 - 1.812q^{-1} + 0.8578q^{-2}}$, $G_2(q)=\frac{1.063q^{-1}}{1 - 1.706q^{-1} + 0.7491q^{-2}}$ and $G_3(q)=1.5G_1(q)$ where $q$ is the forward shift operator, i.e., $qu(t)=u(t+1)$.

    \begin{figure}
  \centering
  \includegraphics[scale=0.46]{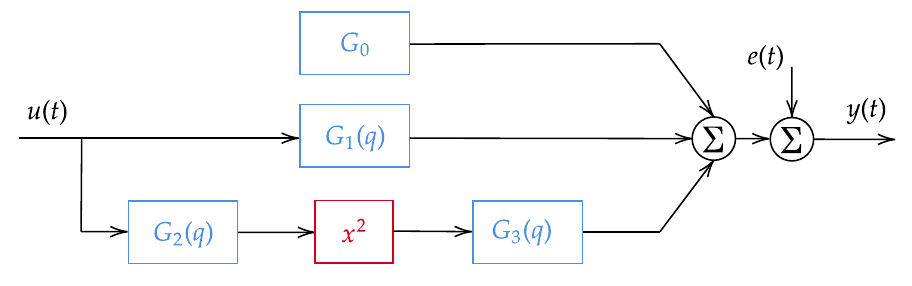}
  \caption{The block-oriented nonlinear system in D1.}
  \label{fig:exp1}
\end{figure}

\item Databank D2 contains $200$ randomly generated WH systems. The static nonlinearity is a polynomial with order $M=2,3$ and the polynomial coefficients are random samples from a uniform distribution on $[-1, 1]$. We have two types of configurations for the linear blocks in particular:
\begin{itemize}
  \item[\textrm{A}] the linear blocks $G_1$ and $G_2$ are randomly generated linear systems of order $30$ with the moduli of the poles in $[0.1,0.9]$
  \item[\textrm{B}] the linear block $G_1$ is a randomly generated linear system of order $15$ and is set to have overdamped dominant dynamics by setting its $5$ poles with the largest moduli real positive and falling in $[0.7,0.8]$ and the moduli of the other poles falling in $[0.1,0.5]$. The linear block $G_2$ is a randomly generated linear system of order $30$ with the moduli of the poles in $[0.1,0.9]$. Moreover, to manifest the correlation between different orders of nonlinearities, the ratios between the variances of the output contributions of the $(m+1)$-th and $m$-th order nonlinearities are kept within $[0.1,10]$.
\end{itemize}

\item Databank D3 contains the Wiener system in \cite[Section 6.1]{LSJ13}. Specifically, the linear block is the $6$-th order system defined by \cite[Eq. (33)]{LSJ13} as follows:
\begin{align}\label{eq:tf_D3}
    G(q) = \frac{\bar{c}_1q^{-1}+\bar{c}_2q^{-2}+\cdots+ \bar{c}_6q^{-6} }{1+\bar{a}_1q^{-1}+\bar{a}_2q^{-2}+\cdots+\bar{a}_6q^{-6}}
\end{align}
with $\bar{a}=[\bar{a}_1,\bar{a}_2,\cdots,\bar{a}_6]$, $\bar{c}=[\bar{c}_1,\bar{c}_2,\cdots,\bar{c}_6]$ and
\[
\begin{split}
    \bar{a}&=[-2.67, 2.96, -2.01, 0.914, -0.181, -0.0102],\\
\bar{c}&=[-0.467, 1.12, -0.925, 0.308, -0.0364, 0.00110].
\end{split}
\]
The static nonlinearity is a saturation function defined by \cite[Eq. (34)]{LSJ13} as follows:
\begin{align}\label{eq:saturation}
\varphi(x)=
\left\{
    \begin{aligned}
    & 1, & x\geq 0.5, \\
    & 2x, & - 0.5 \leq  x < 0.5,\\
    & -1, & x< -0.5.
    \end{aligned}
    \right.
\end{align}

\item Databank D4 contains $50$ randomly generated Wiener systems. The linear block is a randomly generated linear system of order $15$ with the moduli of the poles in $[0.1,0.9]$. The static nonlinearity is a polynomial of order $M=3$ and the polynomial coefficients are random samples from a uniform distribution on $[-1,1]$.
\end{itemize}

Then, we generate the input-output data in D1–D4, where the noise signals are all white Gaussian signals with zero means. The input signals and the signal-to-noise ratios (SNRs) are summarized as follows:
\begin{itemize}
    \item D1: The input signal is a filtered random phase multisine of unit power where the frequency band is set to be $[0,1]$, which is the entire frequency range between 0 and Nyquist frequency, and the number of sine waves is set to be $100$. The SNR is set to be $20$ dB.
    \item D2: The input signal is a white Gaussian signal with zero mean and unit variance. For configuration A in D2, the SNR is set to be $10$ dB. For configuration B in D2, the SNR is set to be $1,5,10$ dB.

    \item D3:  The input signal is a white Gaussian signal with zero mean and unit variance, whose amplitude is within $[-4.2,4]$ and the variance of the noise signal is $0.01$, corresponding to the SNR of $17.23$ dB.

    \item D4:  The input signal is $u(t)=e^{-0.0003t}\cos(0.1t+\pi/3)$. The SNR is set to be $10$ dB.
\end{itemize}

We generate $80$ and $40$ datasets for the single test system in D1 and D3, respectively, and generate one dataset for each randomly generated system in D2 and D4. Specifically, the data in each dataset is generated as follows. We first generate the noiseless data and then
the noiseless data is split into the noiseless training data and the test data. The training data is then formed by adding the noise signals to the output signals in the noiseless training data. The training data is used for model estimation and the test data is used for testing the prediction capability of the model estimate. The number of training data is set to be $N=400$ for D1, D2 and $N=500$ for D3. For D4, each dataset contains $8000$ data points to test the efficiency and we take the first $N=500$ data points as training data. The number of test data is five times that of the training data for D1, D2, D4, and equal to that of the training data for D3.

\subsection{Simulation setup}
The orders of Volterra series are set to be the true orders for D1, D2 and D4, and is set to be $M=9$ for D3. The memory lengths are set to be $n=80$ for D1 and D2, $n=100$ for D3 and $n=50$ for D4. The kernels to be tested for each databank are summarized as follows.

\begin{itemize}
    \item D1: The linear part of the test system, i.e., $G_1(q)$ \cite[Fig. 3]{BMLS17} is modeled based on the DC kernel while the second order nonlinear part, i.e., the WH structure in \cite[Fig. 3]{BMLS17} is modeled first based on the kernel \cite[Eq. (6)]{BMLS17} (denoted by DC2) and then the proposed kernel \eqref{eq:off_diag_kernel_WienerHammersteinType} with $p=q=2$, i.e.,  $\mathcal{K}_{22}(\bm{t}_2,\bm{s}_2)$ (denoted by DCWH) with $\kappa_i(t,s)$, $i=1,2$ being DC kernels.
    \item D2: We test the following kernels: the proposed kernel \eqref{eq:off_diag_kernel_WienerHammersteinType} with a block diagonal structure, i.e., $\mathcal{K}_{mm}(\bm{t}_m,\bm{s}_m)$, for $m=1,\ldots,M$ (denoted by DC-bd), the proposed kernel with off-diagonals, i.e., $\mathcal{K}_{pq}(\bm{t}_p,\bm{s}_q)$ with $\zeta(t)$ defined by 1) and 2) in Corollary \ref{prop:PSDzeta} for $p,q=1,\ldots,M$ (denoted by DC-decay and DC-ob, respectively) where $\kappa_i(t,s)$, $i=1,2$ are DC kernels and the number of orthonormal bases is chosen to be $l=100$ for DC-ob, and the output kernel SED-MPK \cite[Eq. (31)]{LCP21}.

     \item D3: We test the following kernel: the special case \eqref{eq:off_diag_kernel_WienerType} with a block diagonal structure, i.e., $\mathcal{K}^\tw_{mm}(\bm{t}_m,\bm{s}_m)$,  $m=1,\ldots,M$ with $\kappa_1(t,s)$ being a multiple kernel as follows:
    \begin{equation}
    \label{eq:SI2_DC}
        \kappa_1(t,s) = ce^{-\alpha_1 (t+s)}e^{-\beta_1 |t-s|}+ \kappa^{\text{SI2}}(t,s)
    \end{equation}
    with $c\geq 0,\alpha_1>0, \beta_1\geq 0$ being hyper-parameters and $\kappa^{\text{SI2}}(t,s)$ defined by \cite[Eq. (41)]{CHEN18}.

    \item D4: We test the following kernels: the special case of SED-MPK, i.e., \cite[Eq. (36)]{LCP21} (denoted by DC-bd-w), the special case \eqref{eq:off_diag_kernel_WienerType}, i.e., $\mathcal{K}^\tw_{pq}(\bm{t}_p,\bm{s}_q)$ with $\kappa_1(t,s)$ being the DC kernel, $\zeta(t)$ defined by 1) and 2) in Corollary \ref{prop:PSDzeta} (denoted by DC-decay-w and DC-ob-w, respectively) for $p,q=1,\ldots,M$, where the number of orthonormal bases is chosen to be $l=100$ for DC-ob-w.
\end{itemize}

For each dataset in D3, we also compare the proposed KRM with the setup above (denoted by RVS), with the semi-parametric method in \cite{LSJ13} and the prediction error method in \cite{HLW08} (denoted by SEMIP and PEM, respectively), whose specific settings are as follows:
\begin{itemize}
    \item SEMIP+ARD: $20000$ MCMC iterations (out of which $10000$ iterations are considered as burning) are applied. For the linear block, the model order is chosen to be $n_x$ = 10 and the automatic relevance determination (ARD) prior is used. Specifically, the ARD prior does an automatic order selection by overparameterizing the model and promoting sparsity. The static nonlinearity is modeled by a Gaussian process with  a Matérn kernel \cite[Eq. (31)]{LSJ13}.
    \item PEM+HOCV: for the linear block, the model order is chosen by the hold out cross-validation (HOCV). The HOCV is done as follows: models of different orders are estimated using the first 50\% of the training data. Then, the order that minimizes the prediction error on the remaining 50\% of the training data is chosen and the model is re-estimated using the whole training data. The static nonlinearity is modeled by a polynomial with order $9$.
\end{itemize}

For each dataset in D4, we test the efficiency by evaluating the EB cost \eqref{eq:GML} for $625$ times with $1000,2000,\ldots,8000$ data points and then collecting the average computation time over the 50 datasets.

For D1-D3, the performances of different kernels and methods are measured based on the test data by the ``prediction fit'' (PFit) defined as
\begin{equation}
  \label{eq:Fit}
    \text{PFit}=100\% \left(1-\sqrt{\frac{\sum_{i=1}^{N_{\text{te}}}\left(y^{\text{te},0}_i-\hat{y}^{\text{te}}_i\right)^2}{\sum_{i=1}^{N_{\text{te}}}\left(y^{\text{te},0}_i-\bar{y}^{\text{te},0}\right)^2}}\right),
\end{equation}
where $N_{\text{te}}$ is the number of test data, $y^{\text{te},0}_i$ and $\hat{y}^{\text{te}}_i$ are the $i$-th output in the test data and the $i$-th estimated output based on the test data, respectively and $\bar{y}^{\text{te},0}=\frac{1}{N_{\text{te}}}\sum_{i=1}^{N_{\text{te}}}y^{\text{te},0}_i$. Specifically for D3, the identifiability issue is addressed by assuming that the first nonzero element of the true impulse response of the linear block is known. Then the performances of different methods are also measured based on the ``impulse response fit" (GFit) and ``static nonlinearity fit" (NFit) defined by
\begin{subequations}
    \begin{align}
         \begin{split} \label{eq:ImpFit}
    \text{GFit}=100\% \left(1-\sqrt{\frac{\sum_{i=1}^{n}\left(g^{0}_i-\hat{g}_i\right)^2}{\sum_{i=1}^{n}\left(g^{0}_i-\bar{g}^{0}\right)^2}}\right),
  \end{split}\\
  \begin{split}\label{eq:PolyFit}
    \text{NFit}=100\% \left(1-\sqrt{\frac{\sum_{i=1}^{301}\left(y^{\text{nl},0}_i-\hat{y}^{\text{nl}}_i\right)^2}{\sum_{i=1}^{301}\left(y^{\text{nl},0}_i-\bar{y}^{\text{nl},0}\right)^2}}\right),
  \end{split}
      \end{align}
\end{subequations}
respectively. In \eqref{eq:ImpFit}, $n$ is the FIR model order of the linear block, $g^{0}_i$ and $\hat{g}_i$ are the $i$-th elements of the true and estimated impulse responses of the linear block, respectively and $\bar{g}^{0}=\frac{1}{n}\sum_{i=1}^{n}g^{0}_i$. In \eqref{eq:PolyFit}, the static nonlinearity is evaluated at $301$ points on $[-1.5,1.5]$ which are evenly spaced with a step size $0.01$. Here $y^{\text{nl},0}_i$ and $\hat{y}^{\text{nl}}_i$ are the true and estimated static nonlinearities evaluated at the $i$-th point, respectively and  $\bar{y}^{\text{nl},0}=\frac{1}{301}\sum_{i=1}^{301}y^{\text{nl},0}_i$.

\subsection{Simulation results and discussions}
\label{subsec:simu_discussion}
\begin{figure}
  \centering
  \includegraphics[scale=0.39]{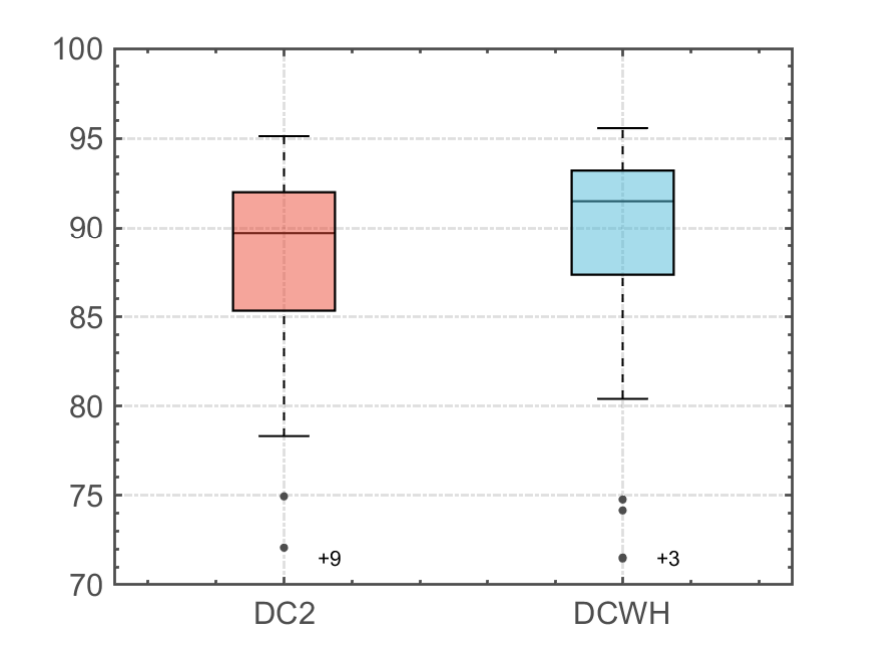}
  \caption{Boxplot of the prediction fits of the estimators based on DC2 and DCWH in Databank D1.}
  \label{fig:exp1_boxplot}
\end{figure}

For Databank D1, the distributions of the prediction fits are shown in Fig. \ref{fig:exp1_boxplot}. The average prediction fits are $82.0578$ and $89.0920$ for DC2 and DCWH, respectively. The result shows a better performance of DCWH than DC2. This is because the proposed kernel DCWH embeds the structural knowledge of the WH system, while DC2 is constructed by the product of two DC kernels after a $45^\circ$ coordinate rotation \cite{BMLS17} and does not encode the correct structural knowledge.

\begin{figure}
  \centering
  \includegraphics[scale=0.46]{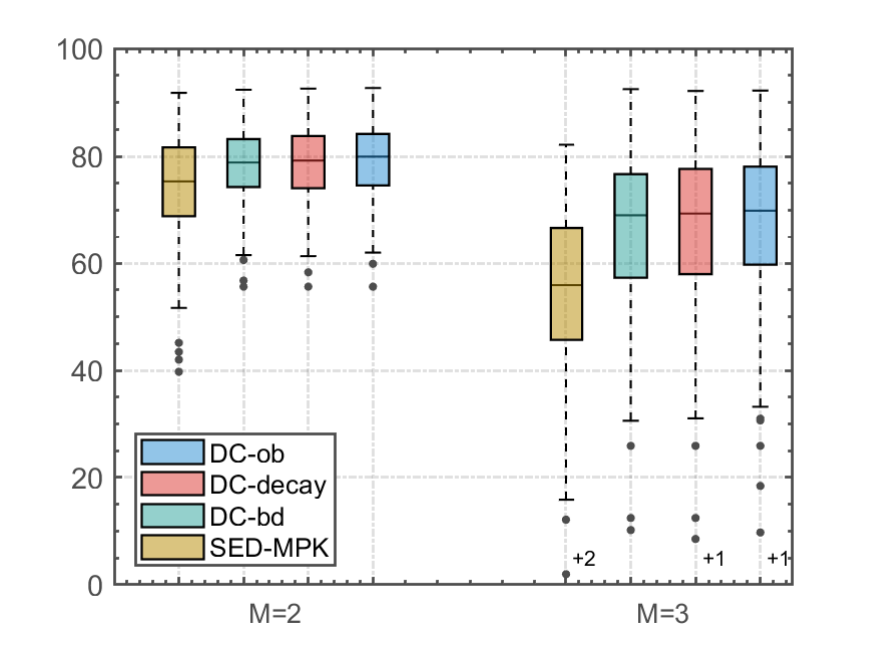}
  \caption{Boxplot of the prediction fits of the estimators based on SED-MPK, DC-bd, DC-decay and DC-ob with $M=2,3$ and SNR$=10$ dB for configuration \textrm{A} in Databank D2.
  }
  \label{fig:exp3_boxplot1}
\end{figure}

\begin{figure}
  \centering
  \includegraphics[scale=0.48]{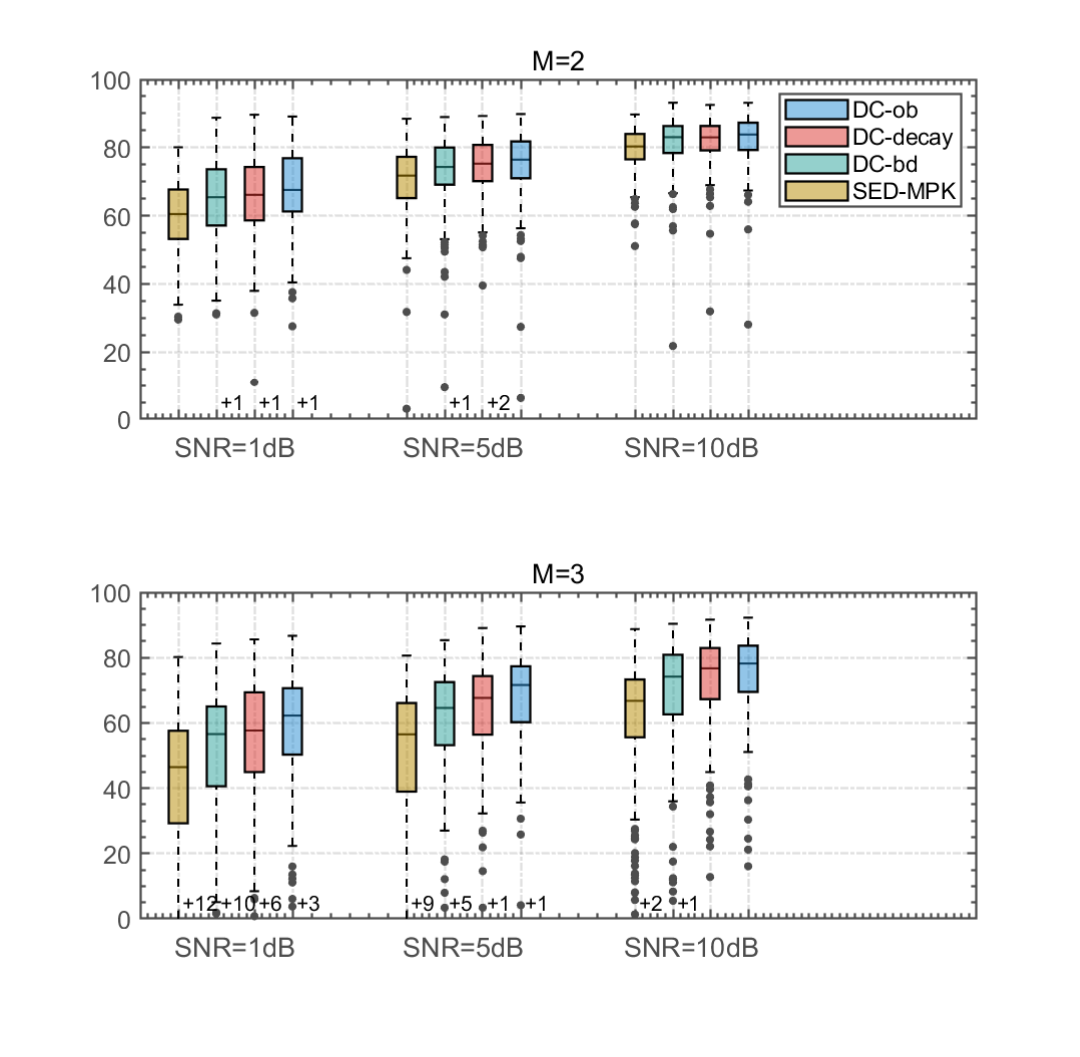}
  \caption{Boxplot of the prediction fits of the estimators based on SED-MPK, DC-bd, DC-decay and DC-ob with $M=2,3$ and SNR$=1,5,10$ dB for configuration \textrm{B} in Databank D2.
}
  \label{fig:exp3_boxplot2}
\end{figure}

For Databank D2, the distributions of the prediction fits for configuration \textrm{A} are shown in Fig. \ref{fig:exp3_boxplot1} and for SED-MPK, DC-bd, DC-decay and DC-ob, the average prediction fits are $74.3923$, $78.6283$, $78.8077$, $79.2181$ and $54.6855$, $66.3461$, $66.7139$, $67.2835$ for $M=2$ and $M=3$, respectively. The distributions of the prediction fits for configuration \textrm{B} are shown in Fig. \ref{fig:exp3_boxplot2} and the average prediction fits are shown in Table \ref{tbl:means_exp3}. For both configurations, the proposed kernels, i.e., DC-bd, DC-decay and DC-ob, outperform SED-MPK. This is because SED-MPK is constructed by the product of DC kernels and it also does not encode the structural knowledge of the WH system. It is also worth to mention that embedded with the correct structural knowledge, the proposed kernels achieve better performances even with less hyper-parameters than SED-MPK. In configuration \textrm{A}, the performances of DC-bd, DC-decay and DC-ob show no obvious differences. In configuration \textrm{B}, the kernels with off-diagonals (DC-ob and DC-decay) outperform the kernel with a block diagonal structure (DC-bd), especially in the cases of low SNRs and/or high orders of nonlinearity. This is because configuration \textrm{B} is designed to manifest the correlation between different orders of nonlinearities and therefore the kernels on the off-diagonal blocks can take effect. In particular, DC-ob outperforms DC-decay in identifying the test systems. This is because $\zeta(t)$ in DC-ob encodes decays with different levels of oscillations but $\zeta(t)$ in DC-decay encodes only an exponential decay, noting that the linear block $G_1$ has the dominant dynamics of overdamped systems. It makes sense that DC-ob is more capable of handling complex dynamics.

\begin{table}[]
  \resizebox{\columnwidth}{!}{%
  \begin{tabular}{llllll}
  \Xhline{0.8pt}
   & SED-MPK &DC-bd & DC-decay   & DC-ob   \\ \Xhline{0.8pt}
  \makecell[l]{$M=2$,\\SNR$=1$ dB}  &59.8827         &64.4003 &65.3021   &67.3597    \\\hline
  \makecell[l]{$M=2$,\\SNR$=5$ dB}  &69.9883           &72.7688 &73.7474  &74.9963       \\\hline
  \makecell[l]{$M=2$,\\SNR$=10$ dB} &79.2944    &81.4889 & 81.8478  &   82.5924        \\ \hline
\end{tabular}}
    \resizebox{\columnwidth}{!}{%
  \begin{tabular}{llllll}
  \Xhline{0.8pt}
   & SED-MPK &DC-bd & DC-decay   & DC-ob   \\ \Xhline{0.8pt}
  \makecell[l]{$M=3$,\\SNR$=1$ dB}  &41.5444       &49.7501 & 53.7171   & 58.2950    \\\hline
  \makecell[l]{$M=3$,\\SNR$=5$ dB}  &50.0526          &59.4642 &64.0468  &67.2144       \\\hline
  \makecell[l]{$M=3$,\\SNR$=10$ dB} &60.3945    &69.0737 &73.1878    &75.0976         \\ \hline
\end{tabular}}
\caption{Average prediction fits of the estimators based on SED-MPK, DC-bd, DC-decay and DC-ob with $M=2,3$ and SNR$=1,5,10$ dB for configuration \textrm{B} in Databank D2.}
\label{tbl:means_exp3}
\end{table}

\begin{figure}
  \centering
  \includegraphics[scale=0.47]{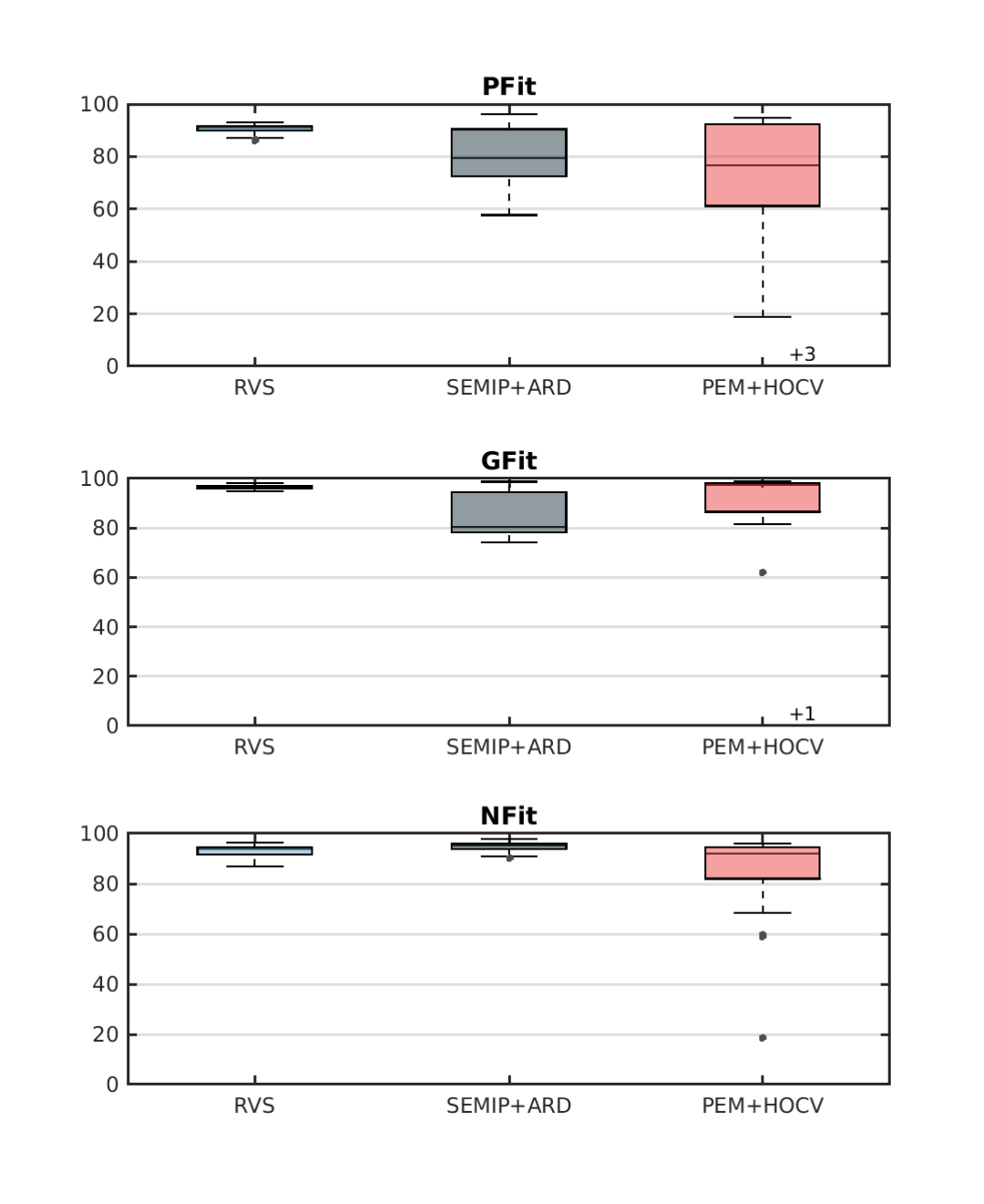}
    \caption{Boxplot of the prediction (top), impulse response (middle) and static nonlinearity (bottom) fits of the estimators based on RVS, SEMIP+ARD and PEM+HOCV in Databank D3.}
    \label{fig:exp_sup_result}
  \end{figure}

For Databank D3, the distributions of the prediction, impulse response and static nonlinearity fits for RVS, SEMIP+ARD and PEM+HOCV are shown in Fig. \ref{fig:exp_sup_result} and their average values are shown in Table \ref{tbl:means_D3}. Specifically, RVS performs the best in terms of accuracy and robustness for the prediction capability. For the impulse response estimation, RVS gives the best GFit in terms of accuracy and robustness and it can be observed that the average GFit of RVS is $14\%$ better than that of SEMIP+ARD. For the static nonlinearity estimation, RVS performs worse than SEMIP+ARD but not very much and better than PEM+HOCV and it can be observed that the average NFit of RVS is only $3\%$ worse than that of SEMIP+ARD.

  \begin{table}[]
    \resizebox{\columnwidth}{!}{%
  \begin{tabular}{ccccc}
  \Xhline{0.8pt}
   & RVS &SEMIP+ARD & PEM+HOCV     \\ \Xhline{0.8pt}
  \makecell[l]{PFit}  &89.8148       &81.7945 & 52.6788      \\\hline
  \makecell[l]{GFit}  &96.4688          &86.7657 &90.4634         \\\hline
  \makecell[l]{NFit} &91.9907    &94.7757 &85.7326            \\ \hline
\end{tabular}}
\caption{Average prediction, impulse response and static nonlinearity fits of the estimators based on RVS, SEMIP+ARD and PEM+HOCV in Databank D3.}
\label{tbl:means_D3}
\end{table}

\begin{Remark}\label{rmk:comparison}
 There exist alternative approaches to Volterra series identification, e.g., the method by using orthonormal basis functions \cite{CFA04} and the sparse methods \cite{KG11}, and it is not clear at the moment whether the proposed method can outperform them in terms of both performance and computational cost.
\end{Remark}

\begin{figure}
  \centering
  \includegraphics[scale=0.49]{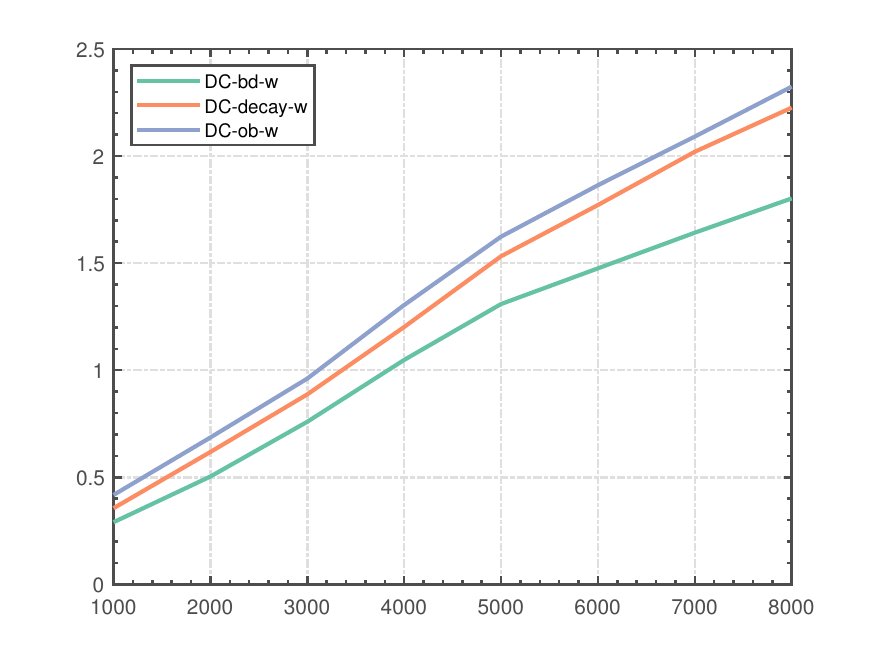}
  \caption{Average computation time (in seconds) of evaluating the EB cost $625$ times over $50$ test systems with number of data $N = 1000, 2000, \ldots , 8000$, based on DC-bd-w, DC-decay-w and DC-ob-w by exploring separability structure in Databank D4.}
  \label{fig:exp4_new}
\end{figure}

For Databank D4, the average computation time of evaluating the EB cost $625$ times over $50$ test systems by exploring the separability structure is shown in Fig. \ref{fig:exp4_new}. The results justify our theory that the computational complexity for computing \eqref{eq:GML} by exploring the separability structure is linear w.r.t. $N$. Moreover, we observe that the computation based on the kernel with a block diagonal structure is more efficient than that based on the kernel with off-diagonals. These observations also justify our theory in Section \ref{sec:RankStructure}. Particularly, the computation by DC-ob-w is more time consuming than DC-decay-w because the construction of the off-diagonals in DC-ob-w involves evaluating $l$ orthonormal bases. We also calculate the prediction fits and it is observed that the error between the prediction fits of the proposed efficient implementation and the standard implementation by direct Cholesky factorization of $Q+\sigma^2 I_N$ is within $10^{-4}$.

\section{Conclusion}
As a major advance of linear system identification in the last decade, the kernel-based regularization method has been employed and achieved success in the identification of nonlinear systems modeled by Volterra series recently. In this paper, we have designed a new kernel that encodes the prior knowledge of Wiener-Hammerstein systems. The designed kernel has nonzero off-diagonal blocks and its structure allows an efficient construction of the output kernel matrix. Efficient implementations have been achieved by exploring the rank structures of the output kernel matrix corresponding to a special case of the kernel, which interestingly encodes the prior knowledge of Wiener systems. Several issues, such as the possible rank structures of the kernel and better designs of the off-diagonal blocks, are to be addressed in future research.

\renewcommand{\thesection}{A}
\setcounter{thm}{0}

\section{Appendix} \label{Appendix}

\subsection{Proof of Proposition \ref{prop:OptimalKernel_offdiagonal_WienerHammersteinType}}
Without loss of generality, here we assume that $p\leq q$ and it follows that
\begin{align*}
        &\mathcal{K}_{pq}^{opt}(\bm{t}_p,\bm{s}_q)=h_p^0(\bm{t}_p)h_q^0(\bm{s}_q)  \notag\\
=& \Big(a_p^0 \sum_{\sigma=0}^{n-1}g_2^0(\sigma)\prod_{i=1}^{p}g_1^0(t_i-\sigma)\Big)\Big(a_q^0 \sum_{\sigma=0}^{n-1}g_2^0(\sigma)\prod_{i=1}^{q}g_1^0(s_i-\sigma)\Big)  \notag\\
=& a_p^0a_q^0 \sum_{\xi_1=0}^{n-1}\sum_{\xi_2=0}^{n-1}g_2^0(\xi_1)g_2^0(\xi_2)  \notag\\&\times\prod_{i=1}^pg_1^0(t_i-\xi_1)g_1^0(s_i-\xi_2)\prod_{i=p+1}^qg^{0}_1(s_i-\xi_2)  \notag\\
        =&a_p^0a_q^0 \sum_{\xi_1=0}^{n-1}\sum_{\xi_2=0}^{n-1}\kappa_2^{\text{opt}}(\xi_1,\xi_2)  \notag\\&\times\prod_{i=1}^p\kappa_1^{\text{opt}}(t_i-\xi_1,s_i-\xi_2)\prod_{i=p+1}^qg^{0}_1(s_i-\xi_2).
\end{align*}
The case with $p>q$ can be obtained in a similar way.

\subsection{Proof of Theorem \ref{Prop:KernelPSD}}
The kernel matrix $P\in\mathbb{R}^{n_\theta\times n_\theta}$ is defined through the kernel $\mathcal{K}_{pq}(\bm{t}_p,\bm{s}_q)$ in \eqref{eq:off_diag_kernel_WienerHammersteinType_general}, and thus can be constructed by
the 2-dimensional convolution between the kernel matrix $K_2\in\mathbb{R}^{n\times n}$ defined through $\kappa_2(t,s)$, and the kernel matrix $P^\tw\in\mathbb{R}^{n_\theta\times n_\theta}$ defined through the kernels $\mathcal{K}_{pq}^\tw(\bm{t}_p,\bm{s}_q)$, $p,q=1,\ldots,M$ in \eqref{eq:off_diag_kernel_WienerHammersteinType_general}. Clearly, if we can show that $P^\tw$ is positive semidefinite, then $P$ will be also positive semidefinite according to the convolution theorem and the Schur product theorem.

Now we show that $P$ is positive semidefinite, that is, for any $v\in\mathbb{R}^{n_\theta}$, $v^TP^\tw v\geq 0$. Let $P^\tw_{pq}\in\mathbb{R}^{n^p\times n^q}$ denote the $\{p,q\}$-th block of $P^\tw$, $p,q=1,\ldots,M$. Then $v$ can be partitioned accordingly by $v=[v_1^T,v_2^T,\cdots,v_M^T]^T$ with $v_m\in\mathbb{R}^{n^m}$. The problem is to prove $v^TP^\tw v = \sum_{p=1}^M\sum_{q=1}^M v_p^TP^\tw_{pq}v_q\geq 0$ and we prove it in four steps.

\emph{Step 1: represent $v_p^TP^\tw_{pq}v_q$ in terms of $K_1$ and $\bm{\zeta}$.}

First, we denote any fixed vectorization of an $m$-way cubical tensor\footnote{An $m$-way cubical tensor is an $\underbrace{n\times n\times \cdots n}_{m \text{ times}}$ multi-dimensional array with $n\in\mathbb{N}$ being its dimension.} $T$ by $\text{Vec}^{(m)}[T]$. Then note that $P^\tw_{pq}$ is constructed by first vectorizing the first $p$ indices and then vectorizing the $(p+1)$-th to the $(p+q)$-th indices of $K_{pq}^\tw$, i.e., $P_{pq}^\tw=\text{Vec}^{(q)}[\text{Vec}^{(p)}[K_{pq}^\tw]]$, where $K_{pq}^\tw$ is the $(p+q)$-way cubical tensor defined by $\mathcal{K}_{pq}^\tw(\bm{t}_p,\bm{s}_q)$ and for $i=1,\ldots,p$ and $j=1,\ldots,q$, $t_i,s_j$ are sampled at $0,1,\ldots,n-1$. Then there also exists an $m$-way cubical tensor $V_m$ corresponding to $v_m\in\mathbb{R}^{n^m}$ such that $\text{Vec}^{(m)}[V_m]=v_m$. For convenience,we denote the $(\tau_1,\cdots,\tau_m)$-th element of $V_m$ by $V_m(\tau_1,\ldots,\tau_m)$ and this convention also holds for vectors, matrices and other tensors in this proof hereafter. Also, we omit $a_1,\ldots,a_M$  since they can be absorbed into the vectors $v_1,v_2,\cdots,v_M$ and thus have no influence on proving the positive semidefiniteness.

Then it holds that for $p\leq q$,
\begin{align}
&v_p^TP_{pq}^\tw v_q= \sum_{i=1}^{n^p}\sum_{j=1}^{n^q} P^{\tw}_{pq}(i,j)v_p(i)v_q(j)  \notag\\
=&\sum_{\tau_1=1}^n\cdots\sum_{\tau_p=1}^n\sum_{\sigma_1=1}^n\cdots\sum_{\sigma_q=1}^nK^\tw_{pq}(\tau_1,\ldots,\tau_p,\sigma_1,\ldots,\sigma_q)&  \notag\\
&\qquad\quad\quad\qquad\times V_p(\tau_1,\ldots,\tau_p)V_q(\sigma_1,\ldots,\sigma_q)  \notag\\
    =&\sum_{\tau_1=1}^n\cdots\sum_{\tau_p=1}^n\sum_{\sigma_1=1}^n\cdots\sum_{\sigma_q=1}^n\prod_{i=1}^pK_1(\tau_i,\sigma_i)\prod_{i=p+1}^q\bm{\zeta}(\sigma_i)&  \notag\\
    &\qquad\quad\quad\qquad\times V_p(\tau_1,\ldots,\tau_p)V_q(\sigma_1,\ldots,\sigma_q)
\end{align}
 and similarly for $p>q$,
\begin{align}
    v_p^T&P^{\tw}_{pq}v_q=  \notag\sum_{\tau_1=1}^n\cdots\sum_{\tau_p=1}^n\sum_{\sigma_1=1}^n\cdots\sum_{\sigma_q=1}^n\prod_{i=1}^qK_1(\tau_i,\sigma_i)\\&\times\prod_{i=q+1}^p\bm{\zeta}(\tau_i) V_p(\tau_1,\ldots,\tau_p)V_q(\sigma_1,\ldots,\sigma_q),
\end{align}
where $K_1\in\mathbb{R}^{n\times n}$ and $\bm{\zeta}\in\mathbb{R}^n$ is the kernel matrix and the vector defined by $K_1(i,j)=\kappa_1(i-1,j-1)$ and $\bm{\zeta}(i)=\zeta(i-1)$, respectively, $i,j=1,\ldots,n$.

Before proceeding to the next step, we define the some notations: $v^{(m)}=[v_m^T,v_{m+1}^T,\cdots,v_M^T]^T$,
\[
    \mathcal{P}^{(m)} = \begin{bmatrix}
    P^{\tw}_{mm}&P^{\tw}_{m(m+1)}&\cdots&P^{\tw}_{mM}\\
      P^{\tw}_{(m+1)m}&P^{\tw}_{(m+1)(m+1)}&\cdots&P^{\tw}_{(m+1)M}\\
      \vdots&\vdots&\ddots&\vdots\\
  P^{\tw}_{Mm}&P^{\tw}_{M(m+1)}&\cdots&P^{\tw}_{MM}
    \end{bmatrix}
\]
and $\chi^{(m)}=(v^{(m)})^T\mathcal{P}^{(m)}v^{(m)}$. It then holds that $\chi^{(1)}=v^TP^{\tw}v$. For $m\leq \min\{p,q\}$, we also define
\begin{align}
  &\Gamma_{pq}^{(m)} = \sum_{\tau_1=1}^n\cdots\sum_{\tau_p=1}^n\sum_{\sigma_1=1}^n\cdots\sum_{\sigma_q=1}^n\prod_{i=1}^{m-1}K_1(\tau_i,\sigma_i)\prod_{j=m}^p\bm{\zeta}(\tau_j)  \notag\\
  &\quad\times\prod_{\ell=m}^q\bm{\zeta}(\sigma_\ell) V_p(\tau_1,\ldots,\tau_p)V_q(\sigma_1,\ldots,\sigma_q)
\end{align}
and $\Gamma^{(m)}=\sum_{p=m}^M\sum_{q=m}^M \Gamma_{pq}^{(m)}$.

\emph{Step 2: prove the recursion $\chi^{(m)} - \chi^{(m+1)}=-\Gamma^{(m+1)}$\\$+\sum_{p=m}^M\sum_{q=m}^M \Gamma_{pq}^{(m+1)}$.}

It holds that
\begin{align}\label{eq:recursion}
  \notag
    &- \Gamma^{(m+1)} + \sum_{p=m}^M\sum_{q=m}^M \Gamma_{pq}^{(m+1)}\\\notag
    =&-\sum_{p=m+1}^M\sum_{q=m+1}^M\Gamma_{pq}^{(m+1)} + \sum_{p=m}^M\sum_{q=m}^M \Gamma_{pq}^{(m+1)}\\\notag
    =&\Gamma_{mm}^{(m+1)}+\sum_{p=m+1}^M \Gamma_{pm}^{(m+1)}+ \sum_{q=m+1}^M \Gamma_{mq}^{(m+1)}\\\notag
    =&\sum_{\tau_1=1}^n\cdots\sum_{\tau_m=1}^n\sum_{\sigma_1=1}^n\cdots\sum_{\sigma_m=1}^n\prod_{i=1}^{m}K_1(\tau_i,\sigma_i)\\\notag
    &\qquad\qquad\times V_m(\tau_1,\ldots,\tau_m)V_m(\sigma_1,\ldots,\sigma_m)\\\notag
    &+\sum_{p=m+1}^M\sum_{\tau_1=1}^n\cdots\sum_{\tau_p=1}^n\sum_{\sigma_1=1}^n\cdots\sum_{\sigma_m=1}^n\prod_{i=1}^{m}K_1(\tau_i,\sigma_i)\\\notag
    &\qquad\times\prod_{j=m+1}^p\bm{\zeta}(\tau_j) V_p(\tau_1,\cdots,\tau_p)V_m(\sigma_1,\cdots,\sigma_m)\\\notag
    &+\sum_{q=m+1}^M\sum_{\tau_1=1}^n\cdots\sum_{\tau_m=1}^n\sum_{\sigma_1=1}^n\cdots\sum_{\sigma_q=1}^n\prod_{i=1}^{m}K_1(\tau_i,\sigma_i)\\\notag
    &\qquad\times \prod_{\ell=m+1}^q\bm{\zeta}(\sigma_\ell)V_m(\tau_1,\ldots,\tau_m)V_q(\sigma_1,\ldots,\sigma_q)
    \\\notag
    =&v_m^TP^\tw_{mm}v_m + \sum_{p=m+1}^M v_p^TP^\tw_{pm}v_m + \sum_{q=m+1}^M v_m^TP^\tw_{mq}v_q\\
=& \chi^{(m)} - \chi^{(m+1)}.
\end{align}
\emph{Step 3: prove the inequality $\chi^{(m)} -\Gamma^{(m)} \geq \chi^{(m+1)} - \Gamma^{(m+1)}$.}

The recursion \eqref{eq:recursion} implies that
$
      \chi^{(m)} -\Gamma^{(m)} =  \chi^{(m+1)} - \Gamma^{(m+1)} + \sum_{p=m}^M\sum_{q=m}^M\Gamma_{pq}^{(m+1)} - \Gamma^{(m)}.
$
To prove the inequality in \emph{Step 3}, here we show that $\sum_{p=m}^M\sum_{q=m}^M\Gamma_{pq}^{(m+1)} - \Gamma^{(m)}\geq 0$. We rewrite $\sum_{p=m}^M\sum_{q=m}^M\Gamma_{pq}^{(m+1)}$ as follows:
\begin{align}
   \notag&\sum_{p=m}^M\sum_{q=m}^M\Gamma_{pq}^{(m+1)}\\\notag
&=\sum_{p=m}^M\sum_{q=m}^M\sum_{\tau_1=1}^n\cdots\sum_{\tau_p=1}^n\sum_{\sigma_1=1}^n\cdots\sum_{\sigma_q=1}^n\prod_{i=1}^{m}K_1(\tau_i,\sigma_i)\\\notag
&\quad\times\prod_{j=m+1}^p\bm{\zeta}(\tau_j)\prod_{\ell=m+1}^q\bm{\zeta}(\sigma_\ell) V_p(\tau_1,\ldots,\tau_p)V_q(\sigma_1,\ldots,\sigma_q)\\\notag
&=\sum_{\tau_1=1}^n\cdots\sum_{\tau_m=1}^n\sum_{\sigma_1=1}^n\cdots\sum_{\sigma_m=1}^n\prod_{i=1}^{m}K_1(\tau_i,\sigma_i)\\&\times W^{(p)}(\tau_1,\ldots,\tau_m)W^{(q)}(\sigma_1,\ldots,\sigma_m)
\end{align}
with
\begin{align*}
  W^{(p)}(\tau_1,&\ldots,\tau_m)\triangleq\\&\sum_{p=m}^M\sum_{\tau_{m+1}=1}^n\cdots\sum_{\tau_p=1}^n\prod_{j=m+1}^p\bm{\zeta}(\tau_j)V_p(\tau_1,\ldots,\tau_p),\\
  W^{(q)}(\sigma_1,&\ldots,\sigma_m)\triangleq\\&\sum_{q=m}^M\sum_{\sigma_{m+1}=1}^n\cdots\sum_{\sigma_q=1}^n\prod_{\ell=m+1}^q\bm{\zeta}(\sigma_\ell)V_q(\sigma_1,\ldots,\sigma_q),
\end{align*}
and rewrite $\Gamma^{(m)}$ as follows:
\begin{align}
  \notag
&\Gamma^{(m)}=
\sum_{p=m}^M\sum_{q=m}^M \Gamma_{pq}^{(m)}\\\notag
=&\sum_{\tau_1=1}^n\cdots\sum_{\tau_{m}=1}^n\sum_{\sigma_1=1}^n\cdots\sum_{\sigma_{m}=1}^n\prod_{i=1}^{m-1}K_1(\tau_i,\sigma_i)\bm{\zeta}(\tau_m)\bm{\zeta}(\sigma_m)\\&\times W^{(p)}(\tau_1,\ldots,\tau_m)  W^{(q)}(\sigma_1,\ldots,\sigma_m).
\end{align}
It follows that
\begin{align}\label{eq:diff}
      &\sum_{p=m}^M\sum_{q=m}^M \Gamma_{pq}^{(m+1)} - \Gamma^{(m)} \notag\\
      =&\sum_{\tau_1=1}^n\cdots\sum_{\tau_{m}=1}^n\sum_{\sigma_1=1}^n\cdots\sum_{\sigma_{m}=1}^n\prod_{i=1}^{m-1}K_1(\tau_i,\sigma_i)\notag\\
      &\times\left(K_1(\tau_m,\sigma_m)-\bm{\zeta}(\tau_m)\bm{\zeta}(\sigma_m)\right)\notag\\
      &\times   W^{(p)}(\tau_1,\ldots,\tau_m)  W^{(q)}(\sigma_1,\ldots,\sigma_m)\geq 0
\end{align}
since $\prod_{i=1}^{m-1}\kappa_1(t_i,s_i)\left(\kappa_1(t_m,s_m)-\zeta(t_m)\zeta(s_m)\right)$ is a positive semidefinite kernel given that $\kappa_1(t,s)-\zeta(t)\zeta(s)$ is a positive semidefinite kernel.

\emph{Step 4: prove that $\chi^{(1)}\geq 0$}

It follows from the inequality in \emph{Step 3} that
$
    \chi^{(1)} -\Gamma^{(1)} \geq \chi^{(2)} -\Gamma^{(2)}\geq \cdots\geq \chi^{(M)} -\Gamma^{(M)}
$
and
\begin{align}
    &\chi^{(M)} -\Gamma^{(M)} = v_M^TP^{\tw}_{MM}v_M-\Gamma_{MM}^{(M)}\notag\\
    =&\sum_{\tau_1=1}^n\cdots\sum_{\tau_M=1}^n\sum_{\sigma_1=1}^n\cdots\sum_{\sigma_M=1}^n\prod_{i=1}^{M-1}K_1(\tau_i,\sigma_i)\notag\\
    &\times\left(K_1(\tau_M,\sigma_M)-\bm{\zeta}(\tau_M)\bm{\zeta}(\sigma_M)\right)\notag\\
    &\times V_M(\tau_1,\ldots,\tau_M)V_M(\sigma_1,\ldots,\sigma_M)\geq 0.
\end{align}
Then we have
\begin{align*}
      \chi^{(1)}\geq \Gamma^{(1)} &= \Big(\sum_{p=1}^M\sum_{\tau_1=1}^n\cdots \sum_{\tau_p=1}^n\prod_{j=1}^p\bm{\zeta}(\tau_j)V_p(\tau_1,\ldots,\tau_p)\Big)^2\notag,\end{align*}
which is nonnegative and thus completes the proof.

 \subsection{Proof of Corollary \ref{prop:PSDzeta}}\label{proof:PSDzeta}
 From Theorem \ref{Prop:KernelPSD}, we only need to prove that $\kappa_1(t,s)-\zeta(t)\zeta(s)$ is a positive semidefinite kernel where $\kappa_1(t,s)=c_1^2e^{-\alpha_1(t+s)}e^{-\beta_1|t-s|}$.

 1) With $\zeta(t)=c_1e^{-(\alpha_1+\beta_1)t}$, it follows that
 \begin{equation*}
   \kappa_1(t,s)-\zeta(t)\zeta(s)=c_1^2e^{-\alpha_1(t+s)}(e^{-\beta_1|t-s|}-e^{-\beta_1(t+s)})
 \end{equation*}
which is the covariance function of the Gaussian process defined by
\begin{equation*}
   g(t) = (2\beta_1)^{\frac{1}{2}}e^{(-\alpha_1-\beta_1)t}\left(z+c_1\int_{\frac{\log 2}{2\beta_1}}^te^{\beta_1\tau}w(\tau)d\tau\right),
 \end{equation*}
 where $z\sim \mathcal{N}(0,\frac{c_1^2}{2\beta_1})$ and $w(t)$ is a white Gaussian noise with zero mean and unit
 variance (independent of $z$).

 2) With $\zeta(t)=c_1\big(\sum_{i=1}^{l}\sqrt{2}\ep_i\psi_i(t)\big)$ where $\ep_i=\frac{1}{(i-\frac{1}{2})^2\pi^2}$ and $\phi_i(t)=\sqrt{2}e^{(-\alpha_1+\beta_1)t}\sin\left((i-\frac{1}{2})\pi e^{-2\beta_1 t}\right)$ and noting that $\{\sqrt{\ep_i}\psi_i\}_{i=1}^\infty$ forms an orthonormal basis of the reproducing kernel Hilbert space induced by the DC kernel $\kappa_1(t,s)$ \cite{CHEN18TAC}, we first consider a general case with $
   \tilde{\zeta}(t)=c_1\big(\sum_{i=1}^{l}\lambda_i\sqrt{\ep_i}\psi_i(t)\big),
$
with $\lambda_i\in\mathbb{R}$. We note that $\zeta(t)$ is a special case of $\tilde{\zeta}(t)$ with $\lambda_i=\sqrt{2\ep_i}$. In what follows, we prove that $\kappa_1(t,s)-\tilde{\zeta}(t)\tilde{\zeta}(s)$ is positive semidefinite if $\|\bm{\lambda}\|_2\leq 1$ with $\bm{\lambda}=[\lambda_1,\lambda_2,\cdots,\lambda_l]^T\in\mathbb{R}^l$.

From \cite{CHEN18TAC}, we have $\kappa_1(t,s)=c_1^2\sum_{i=1}^\infty\ep_i\psi_i(t)\psi_i(s)$ and it follows that
 \begin{align*}
       &\kappa_1(t,s)-\tilde{\zeta}(t)\tilde{\zeta}(s)=c_1^2\sum_{i=1}^\infty\ep_i\psi_i(t)\psi_i(s)\notag\\
       &-c_1^2\big(\sum_{i=1}^{l}\lambda_i\sqrt{\ep_i}\psi_i(t)\big)\big(\sum_{i=1}^{l}\lambda_i\sqrt{\ep_i}\psi_i(s)\big)\notag\\\
       =&c_1^2 \left(\bm{\psi}_{L}(t)^T(I_l-\bm{\lambda}\bm{\lambda}^T)\bm{\psi}_{L}(s)+\bm{\psi}_{R}(t)^T\bm{\psi}_{R}(s)\right)
 \end{align*}
 with $\bm{\psi}_{L}(t) = [\sqrt{\ep_1}\psi_1(t),\cdots,\sqrt{\ep_l}\psi_l(t)]^T\in\mathbb{R}^{l}$, $\bm{\psi}_{R}(t)=[\sqrt{\ep_{l+1}}\psi_{l+1}(t),\sqrt{\ep_{l+2}}\psi_{l+2}(t),\cdots]^T\in\mathbb{R}^{\infty}$. This kernel is positive semidefinite if $I_l-\bm{\lambda}\bm{\lambda}^T$ is a positive semidefinite matrix. We note that the only nonzero eigenvalue of $\bm{\lambda}\bm{\lambda}^T$ is $\|\bm{\lambda}\|_2^2$. Therefore $I_l-\bm{\lambda}\bm{\lambda}^T$ is a positive semidefinite matrix if $\|\bm{\lambda}\|_2\leq 1$.

Now we consider the special case with $\lambda_i=\sqrt{2\ep_i}$ and it follows that
 \begin{align}
       \|\bm{\lambda}\|_2^2 &= 2\sum_{i=1}^l\ep_i = 2\sum_{i=1}^l\frac{1}{(i-\frac{1}{2})^2\pi^2}=\frac{8}{\pi^2}\sum_{i=1}^l\frac{1}{(2i-1)^2}\notag\\
       &<\frac{8}{\pi^2}\sum_{i=1}^\infty\frac{1}{(2i-1)^2}=\frac{8}{\pi^2}\frac{\pi^2}{8}=1.
 \end{align}

\subsection{Proof of Theorem \ref{Thm:OutputKernel}}\label{app:OutputKernel}
Without loss of generality letting $p\leq q$, kernel \eqref{eq:off_diag_kernel_WienerType} is
$
        \mathcal{K}_{pq}^{\tw}(\bm{t}_p,\bm{s}_q)=a_pa_q \prod_{i=1}^p\kappa_1(t_i,s_i)\prod_{i=p+1}^q\zeta(s_i)
$
and we define the linear functional $L^{(m)}_t$ by
\begin{align*}
    \notag L^{(m)}_t[h_m]
  = \sum_{\tau_1=0}^{n-1}\cdots \sum_{\tau_m=0}^{ n-1}h_m(\tau_1,\ldots,\tau_m)\prod_{\tau=\tau_1}^{\tau_m}u(t-\tau).
\end{align*}
For $p\leq q$, the $\{p,q\}$-th order output kernel $\mathcal{Q}_{pq}^{\tw}(t,s) $ for $\mathcal{K}_{pq}^{\tw}(\bm{t}_p,\bm{s}_q)$ is given by
\begin{align}\label{eq:OutputKernel_WienerSystem_pq}
    \notag&\mathcal{Q}_{pq}^{\tw}(t,s) = L_t^{p}[L_s^{q}[\mathcal{K}_{pq}^{\tw}]]\\  \notag
  =&\sum_{\tau_1=0}^{n-1}\cdots\sum_{\tau_p=0}^{n-1}\sum_{\sigma_1=0}^{n-1}\cdots\sum_{\sigma_q=0}^{n-1}\mathcal{K}_{pq}^{\tw}(\tau_1,\ldots,\tau_p,\sigma_1,\ldots,\sigma_q)\\  \notag
  &\times\prod_{i=1}^pu(t-\tau_i)u(s-\sigma_i)\prod_{j=p+1}^qu(s-\sigma_j)\\  \notag
  =&a_pa_q\sum_{\tau_1=0}^{n-1}\cdots\sum_{\tau_p=0}^{n-1}\sum_{\sigma_1=0}^{n-1}\cdots\sum_{\sigma_q=0}^{n-1}\prod_{j=1}^p\kappa_1(\tau_j,\sigma_j)\prod_{j=p+1}^q\zeta(\sigma_j)\\  \notag
  &\qquad\times\prod_{i=1}^pu(t-\tau_i)u(s-\sigma_i)\prod_{j=p+1}^qu(s-\sigma_j)\\  \notag
  =&a_pa_q\Big(\sum_{\tau=0}^{n-1}\sum_{\sigma=0}^{n-1}\kappa_1(\tau,\sigma)u(t-\tau)u(s-\sigma)\Big)^p\\
  &\qquad\quad\qquad\times\Big(\sum_{\sigma=0}^{n-1}\zeta(\sigma)u(s-\sigma)\Big)^{q-p}.
\end{align}
Similarly for $p>q$, we can obtain
\begin{align}\label{eq:OutputKernel_WienerSystem_qp}
  &\mathcal{Q}_{pq}^{\tw}(t,s)   \notag
  =a_pa_q\Big(\sum_{\tau=0}^{n-1}\sum_{\sigma=0}^{n-1}\kappa_1(\tau,\sigma)u(t-\tau)u(s-\sigma)\Big)^q  \notag\\&\qquad\qquad\quad\qquad\times\Big(\sum_{\tau=0}^{n-1}\zeta(\tau)u(t-\tau)\Big)^{p-q}.
\end{align}
Noting that
$
       \mathcal{K}_{pq}(\bm{t}_p,\bm{s}_q)=\sum_{\xi_1=0}^{n-1}\sum_{\xi_2=0}^{n-1}\kappa_2(\xi_1,\xi_2)\mathcal{K}^\tw_{pq}(\bm{t}_p-\xi_1,\bm{s}_q-\xi_2),
$
it follows that
\begin{align*}
      \mathcal{Q}_{pq}(t,s)&=L_t^{(p)}[L_s^{(q)}[\mathcal{K}_{pq}]]  \notag
      \\&=\sum_{\xi_1=0}^{n-1}\sum_{\xi_2=0}^{n-1}\kappa_2(\xi_1,\xi_2)L_{t-\xi_1}^{(p)}[L_{s-\xi_2}^{(q)}[\mathcal{K}^{\tw}_{pq}]]  \notag\\
      &=\sum_{\xi_1=0}^{n-1}\sum_{\xi_2=0}^{n-1}\kappa_2(\xi_1,\xi_2)\mathcal{Q}^w_{pq}(t-\xi_1,s-\xi_2).
\end{align*}
The output kernel $\mathcal{Q}(t,s)$ is then
\begin{align}
  \label{eq:OutputKernel}
        \notag&\mathcal{Q}(t,s) = \sum_{p=1}^M\sum_{q=1}^M \mathcal{Q}_{pq}(t,s)\\   &=\sum_{\xi_1=0}^{n-1}\sum_{\xi_2=0}^{n-1}\kappa_2(\xi_1,\xi_2) \mathcal{Q}^{\tw}(t-\xi_1,s-\xi_2),
\end{align}
where for $p,q=1,\ldots,M$,
\begin{align}\label{eq:OutputKernel_WienerSystem}
      \notag \mathcal{Q}^{\tw}&(t,s)=\sum_{p=1}^M\sum_{q=1}^M \mathcal{Q}^w_{pq}(t,s)
    \\  \notag
    &=\sum_{p>q}a_pa_q\Big(\sum_{\tau=0}^{n-1}\sum_{\sigma=0}^{n-1}\kappa_1(\tau,\sigma)u(t-\tau)u(s-\sigma)\Big)^q\\  \notag&\qquad\qquad\qquad\times\Big(\sum_{\tau=0}^{n-1}\zeta(\tau)u(t-\tau)\Big)^{p-q}\\  \notag
    +&\sum_{p\leq q}a_pa_q\Big(\sum_{\tau=0}^{n-1}\sum_{\sigma=0}^{n-1}\kappa_1(\tau,\sigma)u(t-\tau)u(s-\sigma)\Big)^p\\&\qquad\qquad\times\Big(\sum_{\sigma=0}^{n-1}\zeta(\sigma)u(s-\sigma)\Big)^{q-p}.
\end{align}
It follows from \eqref{eq:OutputKernel} that $Q$ can be calculated by the 2-dimensional convolution of $K_2$ and $Q^{\tw}=\sum_{p=1}^M\sum_{q=1}^MQ^{\tw}_{pq}$ with $[Q^w_{pq}]_{ij}=\mathcal{Q}^w_{pq}(i,j)$. From \eqref{eq:OutputKernel_WienerSystem}, one can check that the output kernel matrix $Q^{\tw}$ can be put into the form in \eqref{eq:OutputKernelMatrix_WienerSystem}.
The calculation of $Q$ includes
\begin{itemize}
    \item matrix multiplication $\Psi K_1\Psi^T$  with $\Psi\in\mathbb{R}^{N\times n}$ and $K_1\in\mathbb{R}^{n\times n}$, which requires $O(N^2n+n^2N)$,
    \item $N\times N$ matrix addition, which requires $O(N^2)$,
    \item $N\times N$ matrix Hadamard powers and products, which requires $O(N^2)$,
    \item 2-dimensional convolution between $N\times N$ and $n\times n$ matrices, which requires $O(Nn\log(Nn))$  by fast Fourier transform.
\end{itemize}
Since $n\leq N$ by assumption, the overall computational complexity of constructing $Q$ is $O(N^3)$. Then the computational bottleneck of \eqref{eq:RLSoutput} and \eqref{eq:GML} is the matrix inversion, which requires $O(N^3)$. Hence the overall computational complexity of \eqref{eq:RLSoutput} and \eqref{eq:GML} is $O(N^3)$.

\subsection{Proof of Theorem \ref{Thm:LowRankOutputKernelMatrix_Full} }\label{app:LowRankOutputKernelMatrix_Full}
Given that the input signal $u(t)$ satisfies \eqref{eq:InputCondition}, we first prove that $\Psi K_1\Psi^T$ is separable with separability rank $r$ as follows. The $(t,s)$-th element of $\Psi K_1\Psi^T$ can be written as
\[
\begin{split}
    [\Psi& K_1\Psi^T]_{t,s} = \sum_{a=0}^{n-1}\sum_{b=0}^{n-1}\kappa(b,a)u(s-a)u(t-b)\\
    & = \sum_{a=0}^{n-1}\sum_{b=0}^{n-1}\kappa(b,a)\Big(\sum_{i=1}^r\pi_i(s)\rho_i(a)\Big)\Big(\sum_{j=1}^r\pi_j(t)\rho_j(b)\Big)\\
    & = \sum_{i=1}^r\sum_{j=1}^r\pi_i(s)\pi_j(t) \sum_{a=0}^{n-1}\sum_{b=0}^{n-1}\kappa(b,a)\rho_i(a)\rho_j(b)
\end{split}
\]
which implies that $\Psi K_1\Psi^T=UV^T$ where $U = [\bm{\pi}_1,\cdots,\bm{\pi}_r]\in\mathbb{R}^{N\times r}$, $V = UH^TK_1H\in\mathbb{R}^{N\times r}$ with $H=[\bm{\rho}_1,\cdots,\bm{\rho}_r]\in\mathbb{R}^{n\times r}$, $\bm{\pi}_i=[\pi(1),\cdots,\pi(N)]^T\in\mathbb{R}^{N}$ and $\bm{\rho}_i=[\rho(0),\cdots,\rho(n-1)]^T\in\mathbb{R}^{n}$, $i=1,\ldots,r$.

Then we prove that $(\Psi K_1\Psi^T)^{\circ m}$ is separable with separability rank $\gamma_m \triangleq \frac{(r+m-1)!}{(r-1)!m!}$. We let $V=[\bar{\bm{\rho}}_1,\cdots,\bar{\bm{\rho}}_r]\in\mathbb{R}^{N\times r}$ with $\bar{\bm{\rho}}_i\in\mathbb{R}^{N}$ being the $i$-th column of $V$, $i=1,\ldots,r$. It follows from the multinomial theorem that
\begin{align}\label{eq:spWienerType}
      &(\Psi K_1\Psi^T)^{\circ m}=(UV^T)^{\circ m}
      =(\sum_{i=1}^r\bm{\pi}_i\bm{\rho}_i^T)^{\circ m}\notag\\
      =&\sum_{b_1+\cdots+b_{r}=m}
      \frac{m!}{b_1!\cdots b_{r}!}(\overset{r}{\underset{i=1}{\circ}}\bm{\pi}_i^{\circ b_i})(\overset{r}{\underset{i=1}{\circ}}\bar{\bm{\rho}}_i^{\circ b_i})^T
\end{align}
where $b_1,\ldots,b_{r}$ are nonnegative integers. Noticing from the multinomial theorem that \eqref{eq:spWienerType} is a summation of $\gamma_m$ outer products of two vectors, $(\Psi K_1\Psi^T)^{\circ m}$ is then separable with separability rank $\gamma_m$ and the columns of its two generators consist of all possible $ \frac{m!}{b_1!\cdots b_{r}!}(\overset{r}{\underset{i=1}{\circ}}\bm{\pi}_i^{\circ b_i})$ and $(\overset{r}{\underset{i=1}{\circ}}\bar{\bm{\rho}}_i^{\circ b_i})$ with $b_1+\cdots+b_{r}=m$, respectively.


We next show that $Q=Q^\tw$ is also separable with separability rank $\gamma\triangleq\gamma_M+2\sum_{m=1}^{M-1}\gamma_m$ and to this goal, we introduce the following lemma, whose proof is straightforward and thus skipped.

\begin{Lemma}\label{lmma:MatrixVecSp}
  Let $Z\in\mathbb{R}^{N\times N}$ be separable with separability rank $d$ and can be represented by $Z=AB^T$ with generators $A=[a_1,\cdots,a_d]\in\mathbb{R}^{N\times d}$, $B=[b_1,\cdots,b_d]\in\mathbb{R}^{N\times d}$ and $a_i,~b_i\in\mathbb{R}^{N}$, $i=1,\ldots,d$. For $x\in\mathbb{R}^{N}$, $Z\circ (xx^T)$ is then separable with separability rank $d$ and can be represented by $Z\circ (xx^T)=\tilde{A}\tilde{B}^T$ with
  $\tilde{A}=A\circ (x\mathbf{e}_d^T)\in\mathbb{R}^{N\times d}$, $\tilde{B}=B\circ (x\mathbf{e}_d^T)\in\mathbb{R}^{N\times d}$ and $\mathbf{e}_d=[1,\cdots,1]^T\in\mathbb{R}^{d}$.
\end{Lemma}

For convenience, we let $X \triangleq \Psi K_1 \Psi^T\in\mathbb{R}^{N\times N}$ and $\bm{\psi} \triangleq \Psi\bm{\zeta}\in\mathbb{R}^{N}$. We define
$
  \bm{\eta}_i \triangleq a_i\mathbf{e}+a_{i+1}\bm{\psi} + \cdots +a_M \bm{\psi}^{\circ M-i}=\sum_{\ell=i}^{M}a_\ell \bm{\psi}^{\circ \ell-i},~
  i=1,2,\ldots,M
$
with $\mathbf{e}=[1,\cdots,1]^T\in\mathbb{R}^{N}$ and the following recursion
\begin{align}
  \label{eq:recurO}
  Q^{(k+1)}
  = &X\circ \Big(\bm{\eta}_{M-k}\bm{\eta}_{M-k}^T\notag\\
  &-(\bm{\psi}\bm{\psi}^T)\circ(\bm{\eta}_{M-k+1}\bm{\eta}_{M-k+1}^T) +Q^{(k)}\Big)
\end{align}
with $Q^{(1)}=a_M^2X$. Following from the recursion \eqref{eq:recurO}, one can check that
\begin{align}\label{eq:OutputKernel_recur}
      &Q=Q^{(M)}=\sum_{m=1}^{M-1}X^{\circ m}\circ (\bm{\eta}_m\bm{\eta}_m^T)\\&-X^{\circ m}\circ \big((\bm{\psi}\circ\bm{\eta}_{m+1})(\bm{\psi}\circ\bm{\eta}_{m+1})^T\big) +a_M^2X^{\circ M}\notag
\end{align}
where $X^{\circ m}\circ (\bm{\eta}_m\bm{\eta}_m^T)$ and $-X^{\circ m}\circ \big((\bm{\psi}\circ\bm{\eta}_{m+1})(\bm{\psi}\circ\bm{\eta}_{m+1})^T\big)$ are separable with separability rank $\gamma_m$ from Lemma \ref{lmma:MatrixVecSp}. Since $a_M^2X^{\circ M}=a_M^2(\Psi K_1 \Psi^T)^{\circ M}$ is separable with separability rank $\gamma_M$, $Q$ is then separable with separability rank $\gamma$.

Noticing that $\gamma$ is in general no less than $r$ and $M$ (see Table \ref{tbl:SpRank} for some examples) and the assumption $n\leq N$, we then show the computational complexity of calculating \eqref{eq:RLSoutput} and \eqref{eq:GML} as follows:
\begin{enumerate}
    \item[1)]  Compute $V=UH^TK_1H$. By matrix-matrix multiplication, the computational complexity is $O(r^2n+n^2r+r^2N)$.

     \item[2)] Form the generators of \eqref{eq:spWienerType} for $m=1,\ldots,M$. In particular, for each $m$, it needs at most $2\gamma_m(m-1)$ Hadamard products between two $N$-dimensional vectors and the overall computational complexity is $O(N(\sum_{m=1}^M(m-1)\gamma_m))$, which is no more than $O(\gamma M N)$.
        \item[3)] Form the generators of \eqref{eq:OutputKernel_recur}. In particular, we first calculate the generators for the first two terms in \eqref{eq:OutputKernel_recur}, i.e., the Hadamard products of the columns of the generators formed in 2) with $\bm{\eta}_m$ and $\bm{\psi}\cdot\bm{\eta}_{m+1}$, for $m=1,\ldots,M-1$,  and then stack them together with the generators of the third term $a_M^2X^{\circ M}=a_M^2(\Psi K_1\Psi^T)^{\circ M}$ as the generators of $Q$. Specifically,  we let $\bar{U}\in\mathbb{R}^{N\times \gamma}$ and $\bar{V}\in\mathbb{R}^{N\times \gamma}$ denote the two generators of $Q$, i.e., $Q=\bar{U}\bar{V}^T$. Then the columns of $\bar{U}$ consist of all possible $\frac{m!}{b_1!\cdots b_{r}!}(\overset{r}{\underset{i=1}{\circ}}\bm{\pi}_i^{\circ b_i})\circ \bm{\eta}_m$, $-\frac{m!}{b_1!\cdots b_{r}!}(\overset{r}{\underset{i=1}{\circ}}\bm{\pi}_i^{\circ b_i})\circ (\psi\circ\bm{\eta}_{m+1})$, for $b_1+\cdots+b_{r}=m$, $m=1,\ldots,M-1$ and all possible $a_M^2\frac{M!}{b_1!\cdots b_{r}!}(\overset{r}{\underset{i=1}{\circ}}\bm{\pi}_i^{\circ b_i})$ for $b_1+\cdots+b_{r}=M$. Similarly, the columns of $\bar{V}$ consist of all possible $(\overset{r}{\underset{i=1}{\circ}}\bar{\bm{\rho}}_i^{\circ b_i})\circ \bm{\eta}_m$, $-(\overset{r}{\underset{i=1}{\circ}}\bar{\bm{\rho}}_i^{\circ b_i})\circ (\psi\circ\bm{\eta}_{m+1})$ for $b_1+\cdots+b_{r}=m$, $m=1,\ldots,M-1$ and all possible $(\overset{r}{\underset{i=1}{\circ}}\bar{\bm{\rho}}_i^{\circ b_i})$ for $b_1+\cdots+b_{r}=M$. With $\frac{m!}{b_1!\cdots b_{r}!}(\overset{r}{\underset{i=1}{\circ}}\bm{\pi}_i^{\circ b_i})$, $-\frac{m!}{b_1!\cdots b_{r}!}(\overset{r}{\underset{i=1}{\circ}}\bm{\pi}_i^{\circ b_i})$, $(\overset{r}{\underset{i=1}{\circ}}\bar{\bm{\rho}}_i^{\circ b_i})$ and $-(\overset{r}{\underset{i=1}{\circ}}\bar{\bm{\rho}}_i^{\circ b_i})$, $m=1,\ldots,M$ already computed in 2), the computational complexity to form the generators of \eqref{eq:spWienerType}  is  $O(\gamma N)$.

    \item[4)] Compute $Y^T(Q+\sigma^2 I_N)^{-1}Y=Y^T(\bar{U}\bar{V}^T+\sigma^2 I_N)^{-1}Y$ in  \eqref{eq:GML} by exploiting the matrix inversion lemma. In particular,
\[
\begin{split}
    Y^T&(\bar{U}\bar{V}^T+\sigma^2 I_N)^{-1}Y\\ &=
\frac{1}{\sigma^2}\|Y\|_2^2-\frac{1}{\sigma^2}Y^T\bar{U}(\sigma^2I_\gamma +\bar{V}^T\bar{U})^{-1}\bar{V}^TY
\end{split}
\]
whose computational complexity is $O(\gamma^2N)$.
    \item[5)] Compute $\log\text{det}(Q+\sigma^2 I_N)$ in \eqref{eq:GML} by exploiting the matrix determinant lemma. In particular,
    \[
\text{det}(Q+\sigma^2 I_N) = \sigma^{2N}\text{det}(I_\gamma+\frac{1}{\sigma^2}\bar{V}^T\bar{U})
    \]
    which can be computed with computational complexity $O(\gamma^2 N)$.
    \item[6)] Compute \eqref{eq:RLSoutput}, i.e.,   $\bm{\phi}P\Phi^T(Q+\sigma^2 I_N)^{-1}Y$. In particular, by exploiting the matrix inversion lemma,
    \[
\begin{split}
\bm{\phi}&P\Phi^T(Q+\sigma^2 I_N)^{-1}Y =
   \bar{u}\bar{V}^T(\bar{U}\bar{V}^T+\sigma^2 I_N)Y  \\
   &= \frac{1}{\sigma^2}\bar{u}\bar{V}^TY-\frac{1}{\sigma^2}\bar{u}\bar{V}^T\bar{U}(\sigma^2 I_\gamma+\bar{V}^T\bar{U})^{-1}\bar{V}^TY
\end{split}
    \]
    where $\bm{\phi}P\Phi^T$ can also be proven to be separable with some generators $\bar{u}\in\mathbb{R}^{1\times \gamma}$ and $\bar{V}\in\mathbb{R}^{N\times \gamma}$, i.e., $\bm{\phi}P\Phi^T=\bar{u}\bar{V}^T$ (we skip the proof since it is similar to the proof for the separability of $Q=\Phi P\Phi^T$). Thus \eqref{eq:RLSoutput} can be computed with computational complexity $O(\gamma^2 N)$.
\end{enumerate}
The overall computational complexity of \eqref{eq:RLSoutput} and \eqref{eq:GML} is
$O(N\gamma^2+n^2r)$.

Finally, if $\kappa_1(t,s)$ is extended-$p$ semiseparable with $p\ll N$, then $K_1H$ in 1) can be computed in $O(rnp)$ by exploring the semiseparable structure of $K_1$, e.g., \cite[Algorithm 4.1]{AC20} and the computational complexity of computing $V$ is $O((r^2+pr)n+r^2N)$. The overall computational complexity of \eqref{eq:RLSoutput} and \eqref{eq:GML} is then reduced to $O(N\gamma^2)$.

\bibliographystyle{unsrt}
\bibliography{database}

\end{document}